# The Two-Nucleon Potential from Chiral Lagrangians

C. Ordóñez

Theory Group, Department of Physics
The University of Texas at Austin, Austin, Texas 78712

and

Department of Physics and Astronomy
Vanderbilt University, Nashville, Tennessee 37203

L. Ray

Department of Physics
The University of Texas at Austin, Austin, Texas 78712

U. van Kolck

Department of Physics, Box 351560
University of Washington, Seattle, Washington 98195

**Abstract**

Chiral symmetry is consistently implemented in the two-nucleon problem at low-energy through the general effective chiral lagrangian. The potential is obtained up to a certain order in chiral perturbation theory both in momentum and coordinate space. Results of a fit to scattering phase shifts and bound state data are presented, where satisfactory agreement is found for laboratory energies up to about 100 MeV.

# 1 Introduction

The problem of deriving the interaction potential between two nucleons continues to be one of the most fundamental problems in nuclear physics. Early field theoretical work in this area [1, 2, 3, 4] encountered many difficulties, mostly due to the non-renormalizability of meson theory. This was followed by more phenomenological approaches which utilized empirical forms for the medium and short-range parts of the interaction potential [5, 6]. During the last two decades a compromise approach has been developed in which meson exchange potentials provide the medium and long-range parts of the nucleon-nucleon (NN) potential while the short-range dynamics is treated phenomenologically [7, 8, 9]. Although the latter approaches have achieved very impressive empirical descriptions of nucleon-nucleon bound state (deuteron) and scattering data the connection between the nucleon-nucleon interaction and the fundamental, underlying dynamics of the strong interaction remains unclear. It is for this reason that the nucleon-nucleon problem continues to be of fundamental interest.

It has been argued [10] that Regge phenomenology can be extended to low-energy nucleon-nucleon scattering with Regge poles leading to a one-boson-exchange (OBE) potential where i) the contributions of meson trajectories (including scalar $\varepsilon$'s) are dominated by the particles with lowest spin which couple to nucleons with a gaussian form factor and ii) gaussian potentials arise from the Pomeron and tensor trajectories. Such a potential in a nonrelativistic expansion has been constructed by the Nijmegen group [7] and fits data very well. However, Regge cuts are simply neglected. The Bonn group [8] made a serious attempt to include multi-boson exchange in the framework of old-fashioned perturbation theory. In addition to the OBE of known mesons they included the following: $2\pi$ and $\pi\rho$ exchange with both nucleons and $\Delta$ isobars in intermediate states, "correlated" two-pion exchange in the form of a $\sigma'$ scalar meson, $\pi\sigma_{OBE}$ exchange (with $\sigma_{OBE}$ an approximation to $2\pi$, $\sigma'$ and $\pi\rho$ exchanges), and $\pi\omega$ exchange. Agreement with data is quite good.

Nevertheless, the justification for such approaches in terms of quantum chromodynamics (QCD) remains mysterious. In particular, it is not clear how to consistently deal with the exchange of mesons which have masses of the order of the typical inverse hadronic radius set by the QCD scale $\Lambda_{QCD}$. This has led a number of researchers [11] to attempt derivations of nucleon-nucleon scattering from quark models (either constituent or bag) formulated in terms of some effective degrees of freedom which carry the same quantum numbers as the current quarks and gluons. Although such models are not derived from QCD either, they usually have only a few parameters, most of which are fixed by fitting one-nucleon properties. Generally these models produce adequate short-range interactions [11], but the long range potential continues to be formulated in terms of pion exchange.

It seems natural, therefore, to start a treatment of the nuclear force problem by recognizing the unique role played by the pion. Although we are largely ignorant of the non-perturbative dynamics of QCD at low energies, we know there exists an approximate chiral symmetry which is broken by the vacuum. This symmetry restricts the form of the allowed interactions of pions among themselves and with other particles. Consequences of approximate chiral symmetry are i) the small mass of the pion relative to the QCD scale, $\Lambda_{QCD}$, and its subsequent long-range contribution to the NN potential and ii) theorems relating processes involving different numbers of pions which yield some predictive power. The pion is indeed the most important character, besides the nucleon, in the nuclear physics drama.

The distinguished status of the pion in determining the NN interaction has, of course, been emphasized before, particularly by the Stony Brook and Paris groups [9]. The coordinate space potential developed by the latter contains: i) a long range, "theoretical" part constructed through unitarity, analyticity and crossing relations from $\pi\pi$ and $\pi N$ phase shifts, which includes one, two (continuum plus $\rho$, $\varepsilon$) and partially three pion exchanges (in the form of $\omega$) and ii) short range, purely phenomenological spin and isospin dependent parts. Both groups evolved from this model-independent but parameter-crowded approach to the other extreme, the two-parameter Skyrme model. Semi-quantitative success resulted, except for the lack of a central, intermediate range attraction [12]. (For a review of further developments, see [13].)

What is fundamentally new in the present approach [14, 15] is the development of the NN potential within the framework of the general effective chiral lagrangian. By considering the most general lagrangian which involves the pion and the nucleon, and transforms under chiral symmetry as the QCD lagrangian, we divide the NN problem into two parts. The first task concerns QCD and its reformulation in terms of the relevant, low-energy degrees of freedom. The resulting theory must have the form of the general chiral lagrangian (because the latter contains *all* the interactions with the correct symmetry), where the coupling constants are, in principle, known functions of fundamental quantities like $\Lambda_{QCD}$ and the quark masses. In other words, the dynamics of QCD is buried in the couplings of the effective chiral lagrangian. Since different models based on QCD represent different attempts to capture the essence of this underlying dynamics, they will generally differ in the strengths of the low-energy parameters. The second part of the problem is to relate the parameters of the effective chiral lagrangian to the measured, low-energy NN scattering data and deuteron properties.

Clearly, we do not attempt to "solve" QCD here, but instead concentrate on the second task described in the preceeding paragraph. We start with the general chiral lagrangian with undetermined coefficients. Because chiral symmetry is manifest (contrary to most meson-exchange models—e.g. [7, 8]), our approach is *a priori* compatible both with QCD and with all known low-energy phenomenology, including $\pi\pi$, $\pi N$, and $\gamma N$ scattering, meson-exchange currents, etc. Our scheme is model independent in the sense that we do not adopt either a massive meson exchange picture or a particular quark model. When a systematic analysis based on a chiral lagrangian is carried out for such processes as $\pi N$ scattering, a number of the unknown coefficients in our model can be determined independently of the NN data. In the meantime we keep these parameters free in the NN data fitting procedure.

We do have to make one assumption, that of naturalness, which requires that the parameters be consistent with naive dimensional analysis. With this one assumption a perturbative treatment of the nuclear potential can be developed that is lacking in other approaches. Here the perturbative expansion is in powers of momentum divided by a typical QCD mass scale. Up to a given order of expansion the effective chiral lagrangian specifies precisely the terms which appear in the NN potential. Of course, there is no guarantee that the resulting potential will be sufficient to describe the data. If a good overall description of the data results, it means that the perturbation expansion was carried out to the order the precision of the data requires. If, on the other hand, an important phenomenological ingredient (*e.g.* scalar isoscalar attraction) is missing, then this might indicate that a certain operator or diagram is more important than naively expected. This in turn would be indicative of some characteristic dynamic mechanism, and we would be learning something about QCD.

We emphasize that our aim here is not to obtain better fits to the nucleon-nucleon data than the already excellent fits achieved with meson-exchange potentials. We do intend for this approach to help establish a bridge between QCD and nuclear physics and to provide a sound model of the nucleon-nucleon potential whose off shell structure is fixed and which may be used for calculating other nuclear processes. In short, the general chiral lagrangian is a useful way to parametrize both our ignorance of QCD and our knowledge of nuclear physics.

General ingredients and properties of effective chiral lagrangians for nuclear physics applications are discussed in Sec. 2; the effective chiral lagrangian expansion used here is presented in Sec. 3. The two-nucleon potential is derived to a certain order in chiral perturbation theory in momentum space in Sec. 4 and in Sec. 5 is transformed into coordinate space, using a momentum space gaussian cut-off. The special techniques required to calculate NN scattering and bound state properties with the present coordinate space potential are discussed in Sec. 6 and the results of fitting the nucleon-nucleon scattering and bound state data are presented in Sec. 7. Conclusions are given in Sec. 8. Finally, many details are deferred to the Appendices. An initial report of these results was presented in Refs. [14, 15].

## 2   Power Counting

In this work the low energy NN potential is expanded in powers of momentum divided by a QCD mass scale. Typical three-momenta $Q$ exchanged in nuclei can be estimated as the inverse of the rms electromagnetic radius $\langle r_{ch}^2 \rangle^{1/2}$ of a light nucleus. For example, for the triton with $\langle r_{ch}^2 \rangle^{1/2} \simeq 1.75$fm we find that $Q \sim m_\pi$, the pion mass. In QCD the coupling becomes strong and is dominated by non-perturbative effects below a momentum scale $M$ that is roughly given by a typical hadronic mass, $\sim 1$ GeV. Whenever we face such a two-scale problem it is useful to separate the corresponding physics by considering an effective, low-energy theory which involves only the relevant degrees of freedom, all with small three-momenta $Q$. Such theories can be formulated with a lagrangian that is local (in the sense that it involves only operators containing fields at the same spacetime point) and shares the symmetries of the underlying theory, in this case QCD. The dynamical information for modes with momenta $\gtrsim M$ is contained in an infinite set of parameters.

What then are the relevant degrees of freedom in the case of low energy nuclear physics? Unlike the situation at high energies where quark and gluon degrees of freedom are indirectly manifest in the data (*e.g.* jets, deep inelastic scattering, quarkonium production, etc.), low energy nuclear physics does not reveal this underlying QCD structure in any obvious way. Therefore the relevant fields for this study should represent mesons and baryons. Clearly, the lightest stable particles in each sector should be included. The pion $\boldsymbol{\pi}$ has a mass that is small compared to $M$, and its pseudo-Goldstone boson nature makes it a fundamental ingredient. The nucleon $N$ has a mass $m_N$ which is not small but because protons and neutrons comprise the principle constituents of nuclei they must be included. (The explicit appearance of the nucleon mass $m_N$ in the effective theory requires care as has been discussed previously [16, 17]). The effects of higher mass meson and baryon states will generally be suppressed by the inverse of the meson masses or by the inverse of the mass differences between the baryons and the nucleon. We retain only those mass states for which this factor is much larger than $\sim 1/M$. In the meson sector, this implies that we do not explicitly keep the $\rho$, $\omega$, etc., whose masses are $\gtrsim 5.5 m_\pi$ which are closer to $M$ than to $m_\pi$. In the baryon sector we retain only the $\Delta$ isobar which has a mass $m_\Delta \sim m_N + 2m_\pi$, but do not include the $N^*$ with mass $m_{N^*} \sim m_N + 3.5 m_\pi$ nor any other higher mass baryon state. The contributions of these additional fields could be included in a similar way as is done for the $\Delta$. The other octet pseudo-Goldstone bosons and the hyperons are also omitted. For simplicity we consider only $SU(2) \times SU(2)$, however our treatment can be readily extended to $SU(3) \times SU(3)$ to encompass hypernuclear physics.

The requirement that the low energy lagrangian incorporates the symmetries of QCD restricts the form of possible interactions involving $\boldsymbol{\pi}$, $N$ and $\Delta$, but we are still left with an infinite set of interactions $i$ with coupling constants $g_i$, which differ in the number of derivatives or powers of pion mass $d_i$, fermion fields $f_i$, etc. If we knew how to solve QCD at low energies, we could calculate these coupling constants directly. Since there is no *a priori* reason for the couplings in the effective chiral lagrangian to be small, no *a priori* perturbation expansion for the infinite set of interactions can be formulated.

We can proceed only by making an assumption of naturalness which means that when a coupling constant $g_i$ of mass dimension $-\delta_i$ is expressed as $g_i = \tilde{g}_i M^{-\delta_i}$, the dimensionless coupling constant $\tilde{g}_i$ will be of order unity. Of course this might not be true for all the couplings and this will become apparent through phenomenological data analysis. If a coupling constant is found to be anomalously large or small, it may require special treatment at low energies, but this may also indicate a particular dynamical or symmetry effect at the level of QCD.

We now have a natural expansion parameter $\frac{Q}{M} \sim \frac{m_\pi}{M}$, the contribution of any diagram being characterized by the power $\nu$ of the soft momentum $Q$. We organize our perturbation expansion by counting powers of $Q$ in the same way that is done to get the superficial degree of divergence of a graph, where special care is taken with baryons due to explicit factors which contain their large masses. In the present effective theory it is assumed that all three-momenta $Q \ll m_N$; nucleons and $\Delta$'s are therefore nonrelativistic [1].

The first task is to organize the expansion in such a way as to eliminate time-derivatives of the fermions in interaction terms, since they would contribute large factors. This has been done by redefining the fermion fields in terms of velocity eigenstates [19], but also more simply by directly replacing the time-derivatives of fermion fields using the equations of motion for the fermions [16, 17]. In so doing we generate interaction terms that have already been accounted for, which simply result in a redefinition of existing coefficients.

The second task is to distinguish between so-called reducible and irreducible diagrams. Reducible diagrams are those which can be separated into two parts by cutting through an intermediate state which contains only the initial or final particles. This type of intermediate state produces infrared divergences in the limit when the baryon kinetic energy is ignored; when it is not, a small recoil energy denominator results which makes the overall diagram bigger than expected by a factor $\frac{m_N}{Q} \gg 1$. The contributions of these reducible diagrams are automatically included by solving the Lippmann-Schwinger or Schrödinger (in the nonrelativistic limit) equations of motion.

The simplest way to isolate these two types of diagrams is to work in the framework of old-fashioned, time-ordered perturbation theory. Irreducible diagrams are those that contain only intermediate states with energies that differ from the initial energy by an amount $O(Q)$. For an irreducible diagram with $V_i$ vertices of type $i$, $L$ loops, $C$ separately connected pieces and $E_f = 2A$ external fermion lines, the power of $Q$ can be conveniently written as

$$\nu = 4 - A + 2L - 2C + \sum_i V_i \Delta_i \qquad (1)$$

where

$$\Delta_i = d_i + \frac{f_i}{2} - 2 \qquad (2)$$

is called the index of vertex $i$. Any reducible diagram can be constructed from irreducible diagrams by connecting the latter with intermediate states with energies that differ from the initial energy by an amount $O(Q^2/m_N)$ or smaller.

---

[1]Since we do not know *a priori* what the scale $M$ is exactly, it is not clear how relativistic corrections (which are suppressed by $1/m_N$) compare to $1/M$ corrections. A rough idea of their relative importance can be obtained from the following naive dimensional argument. The nucleon-nucleon potential in momentum space can be written as $V(p, p') = \alpha I(p, p')$ where $I(p, p')$ is some dimensionless function of the initial and final c.m. momenta $p$ and $p'$, respectively, and $\alpha \sim 2\pi^2/M^2$ if we count powers of 2 and $\pi$ *à la* [18]. Substituting this in the Lippmann-Schwinger equation we obtain an expansion in $\alpha Q m_N/2\pi^2 \sim Q m_N/M^2$. A shallow bound-state indicates that this series barely diverges, so we estimate that $M^2 \sim Q m_N$. This estimate is admittedly crude and it is not crucial for our approach but it suggests that relativistic corrections $O(\frac{Q}{m_N})$ are $O(\frac{Q^2}{M^2})$. If $M$ is actually larger, it only indicates that relativistic corrections are relatively a little larger than assumed here.

Here we deal with diagrams involving only two external nucleons. Irreducible diagrams are then two-nucleon irreducible; any intermediate state contains at least one pion or isobar. The two-nucleon potential is defined as the sum of such irreducible diagrams, their contributions being ordered by Eq. (1). The full NN scattering amplitude is evaluated by iterating the nuclear potential in the Lippmann-Schwinger equation, or equivalently, by solving (numerically) the corresponding Schrödinger equation.

# 3 Effective Chiral Lagrangian

In order to construct a perturbative expansion in $Q/M$, Eq. (1) requires $\Delta_i \geq 0$, for in this case there is a lower bound for $\nu$ corresponding to diagrams with the maximum number of separately connected pieces, no loops and all vertices having $\Delta_i = 0$. Corrections with higher $\nu$ are obtained by inserting loops and interactions with $\Delta_i > 0$, and decreasing the number of connected pieces. We will show that chiral symmetry requires

$$\Delta_i \geq 0. \tag{3}$$

Here, for simplicity, we work with QCD with only two light flavors $u$ and $d$ with masses $m_u$ and $m_d$, but it is straightforward to include the strange quark. In the limit of vanishing quark masses there is an $SU(2) \times SU(2) \sim SO(4)$ symmetry which is spontaneously broken to $SU(2) \sim SO(3)$. As a result, there exist Goldstone bosons whose fields live in the three-sphere $S^3 \sim SO(4)/SO(3)$, with a diameter that turns out to be the pion decay constant $F_\pi \simeq 190 \text{MeV}$. Following Weinberg [16, 17] we use stereographic coordinates $\boldsymbol{\pi}$; the covariant derivative is then

$$\boldsymbol{D_\mu} = \frac{1}{1 + \boldsymbol{\pi}^2/F_\pi^2} \frac{\partial_\mu \boldsymbol{\pi}}{F_\pi} \equiv D^{-1} \frac{\partial_\mu \boldsymbol{\pi}}{F_\pi}. \tag{4}$$

The baryons considered here provide the 1/2 and 3/2 representations of the spin and isospin $SU(2)$ groups. A nucleon $N$ (isobar $\Delta$) is described by a Pauli spinor (a 4-component spinor) in both spin and isospin spaces, the respective generators being denoted by $\frac{1}{2}\vec{\sigma}(\frac{1}{2}\vec{\sigma}^{(3/2)})$ and $\boldsymbol{t}(\boldsymbol{t}^{(3/2)})$. There are also, of course, $2 \times 4$ transition operators $\frac{1}{2}\vec{S}$ and $\boldsymbol{T}$, satisfying

$$S_i S_j^+ = \frac{1}{3}(2\delta_{ij} - i\varepsilon_{ijk}\sigma_k) \tag{5}$$

$$T_a T_b^+ = \frac{1}{6}(\delta_{ab} - i\varepsilon_{abc}t_c), \tag{6}$$

which allow us to couple $N$ and $\Delta$ in bilinear terms with spin and isospin transfer 1, respectively.

The effective chiral lagrangian is constructed out of the fields $\boldsymbol{D}_\mu$, $N$ and $\Delta$ and their covariant derivatives,

$$\mathcal{D}_\mu \boldsymbol{D}_\nu = \partial_\mu \boldsymbol{D}_\nu + i\boldsymbol{E}_\mu \times \boldsymbol{D}_\nu \tag{7}$$

$$\mathcal{D}_\mu N = (\partial_\mu + \boldsymbol{t} \cdot \boldsymbol{E}_\mu)N \tag{8}$$

$$\mathcal{D}_\mu \Delta = (\partial_\mu + \boldsymbol{t}^{(3/2)} \cdot \boldsymbol{E}_\mu)\Delta, \tag{9}$$

where

$$\boldsymbol{E}_\mu \equiv \frac{2i}{F_\pi}\boldsymbol{\pi} \times \boldsymbol{D}_\mu. \tag{10}$$

This is done by considering all possible isoscalar terms and imposing the discrete spacetime symmetries of QCD, parity and time-reversal.

That is not all though, because the quark masses break $SO(4)$ explicitly. The symmetry breaking terms can be written as a linear combination of the fourth component of a chiral four-vector and the third component of another four-vector, with coefficients $\frac{1}{2}(m_u + m_d)$ and $\frac{1}{2}(m_u - m_d)$, respectively. We account for this explicit symmetry breaking by including in the chiral lagrangian all the terms constructed out of $\boldsymbol{\pi}$, $N$ and $\Delta$ that transform under $SO(4)$ in the same way. Their coefficients will then be proportional to powers of these combinations of quark masses. That is the way the pion mass arises, $m_\pi^2 \propto (m_u + m_d)$, so each power of $m_u + m_d$ will count as $Q^2$. For simplicity we neglect isospin breaking terms proportional to $(m_u - m_d)$. When the latter are included along similar lines we begin to understand why isospin violating effects are so feeble in most nuclear phenomena [20]. Appendix A presents further details regarding the transformation properties of the field representation used here.

By writing operators that are chiral invariant or that break chiral invariance proportional to the quark mass term, we immediately see that all interaction terms have $\Delta_i \geq 0$; operators involving only pions have at least two derivatives or two powers of $m_\pi$ and nucleon bilinears have at least one derivative. *Chiral symmetry therefore guarantees a natural perturbative low-energy theory.*

The index of interaction $\Delta_i$ provides a useful ordering scheme for the chiral lagrangian. Below we denote by $\mathcal{L}^{(n)}$, referred to as the $n$-th order lagrangian, the collection of terms with indices $\Delta_i = n$. We explicitly show only those terms relevant for our application. Since we evaluate diagrams only up to one-loop, interaction operators with more pion fields or isobars than those exhibited below do not contribute to this potential, although they are there in general, in many cases to assure chiral invariance. Note also that we eliminate some redundant terms by integrating by parts, by using the equations of motion (e.g. to eliminate nucleon time-derivatives), and by applying Fierz reordering [21].

The lowest order lagrangian is

$$\begin{aligned}\mathcal{L}^{(0)} = & -\frac{1}{2}D^{-2}((\vec{\nabla}\boldsymbol{\pi})^2 - \dot{\boldsymbol{\pi}}^2) - \frac{1}{2}D^{-1}m_\pi^2\boldsymbol{\pi}^2 \\ & + \bar{N}[i\partial_0 - 2D^{-1}F_\pi^{-2}\boldsymbol{t} \cdot (\boldsymbol{\pi} \times \dot{\boldsymbol{\pi}}) - m_N]N \\ & - 2D^{-1}F_\pi^{-1}g_A \bar{N}(\boldsymbol{t} \cdot \vec{\sigma} \cdot \vec{\nabla}\boldsymbol{\pi})N\end{aligned}$$

$$-\frac{1}{2}C_S \bar{N}N\bar{N}N - \frac{1}{2}C_T \bar{N}\vec{\sigma}N \cdot \bar{N}\vec{\sigma}N$$
$$+\bar{\Delta}[i\partial_0 - 2D^{-1}F_\pi^{-2}\boldsymbol{t}^{(3/2)} \cdot (\boldsymbol{\pi} \times \dot{\boldsymbol{\pi}}) - m_\Delta]\Delta$$
$$-2D^{-1}F_\pi^{-1}h_A[\bar{N}\boldsymbol{T} \cdot (\vec{S} \cdot \vec{\nabla}\boldsymbol{\pi})\Delta + h.c.]$$
$$+\ldots \qquad (11)$$

where $g_A$ is the axial vector coupling of the nucleon, $h_A$ is the $\Delta N\pi$ coupling, $C_S$ and $C_T$ are the parameters first introduced by Weinberg [16, 17], and as usual we work in units where $\hbar c = 1$.

In this work we will also employ terms with more derivatives and powers of $\boldsymbol{\pi}$. The first-order lagrangian is

$$\begin{aligned}\mathcal{L}^{(1)} &= -\frac{B_1}{F_\pi^2}D^{-2}\bar{N}N[(\vec{\nabla}\boldsymbol{\pi})^2 - \dot{\boldsymbol{\pi}}^2] \\ &\quad -\frac{B_2}{F_\pi^2}D^{-2}\varepsilon_{ijk}\varepsilon_{abc}\bar{N}\sigma_k t_c N \partial_i \pi_a \partial_j \pi_b \\ &\quad -\frac{B_3}{F_\pi^2}m_\pi^2 D^{-1}\bar{N}N\boldsymbol{\pi}^2 \\ &\quad +\ldots \end{aligned} \qquad (12)$$

where the $B_i$'s are coefficients of order $O(1/M)$; in particular, the last interaction term proportional to $B_3$ contributes to a scalar-isoscalar term similar to the $\sigma$ term in meson exchange potentials. The second-order lagrangian is

$$\begin{aligned}\mathcal{L}^{(2)} &= \frac{1}{2m_N}\bar{N}\vec{\nabla}^2 N - \frac{A'_1}{F_\pi}[\bar{N}(\boldsymbol{t}\cdot\vec{\sigma}\cdot\vec{\nabla}\boldsymbol{\pi})\vec{\nabla}^2 N + \overline{\vec{\nabla}^2 N}(\boldsymbol{t}\cdot\vec{\sigma}\cdot\vec{\nabla}\boldsymbol{\pi})N] \\ &\quad -\frac{A'_2}{F_\pi}\overline{\vec{\nabla}N}(\boldsymbol{t}\cdot\vec{\sigma}\cdot\vec{\nabla}\boldsymbol{\pi})\cdot\vec{\nabla}N \\ &\quad -C'_1[(\bar{N}\vec{\nabla}N)^2 + (\overline{\vec{\nabla}N}N)^2] - C'_2(\bar{N}\vec{\nabla}N)\cdot(\overline{\vec{\nabla}N}N) \\ &\quad -C'_3\bar{N}N[\bar{N}\vec{\nabla}^2 N + \overline{\vec{\nabla}^2 N}N] \\ &\quad -iC'_4[\bar{N}\vec{\nabla}N\cdot(\overline{\vec{\nabla}N}\times\vec{\sigma}N) + (\overline{\vec{\nabla}N})N\cdot(\bar{N}\vec{\sigma}\times\vec{\nabla}N)] \\ &\quad -iC'_5\bar{N}N(\overline{\vec{\nabla}N}\cdot\vec{\sigma}\times\vec{\nabla}N) - iC'_6(\bar{N}\vec{\sigma}N)\cdot(\overline{\vec{\nabla}N}\times\vec{\nabla}N) \\ &\quad -(C'_7\delta_{ik}\delta_{jl} + C'_8\delta_{il}\delta_{kj} + C'_9\delta_{ij}\delta_{kl})\times \\ &\qquad\qquad [\bar{N}\sigma_k\partial_i N\bar{N}\sigma_l\partial_j N + \overline{\partial_i N}\sigma_k N\overline{\partial_j N}\sigma_l N] \\ &\quad -(C'_{10}\delta_{ik}\delta_{jl} + C'_{11}\delta_{il}\delta_{kj} + C'_{12}\delta_{ij}\delta_{kl})\bar{N}\sigma_k\partial_i N\overline{\partial_j N}\sigma_l N \\ &\quad -(\frac{1}{2}C'_{13}(\delta_{ik}\delta_{jl}+\delta_{il}\delta_{kj}) + C'_{14}\delta_{ij}\delta_{kl})\times \\ &\qquad\qquad [\overline{\partial_i N}\sigma_k\partial_j N + \overline{\partial_j N}\sigma_k\partial_i N]\bar{N}\sigma_l N \\ &\quad +\ldots \end{aligned} \qquad (13)$$

where the $A'_i$ and $C'_i$ are additional undetermined coefficients of order $O(1/M^2)$.

Using this expansion for the lagrangian and the rules for diagrams in time-ordered perturbation theory, it is straightforward to construct the interaction potential. Because we eliminated time derivatives in all interaction terms but four (those that come together with the pion and fermion kinetic terms in $\mathcal{L}^{(0)}$, and the $B_1$ term in $\mathcal{L}^{(1)}$), and because each of these four terms involves at least two pion fields, the interaction hamiltonian is just $(-1)$ times the interaction lagrangian, up to interactions with more pion fields that do not contribute to the order we are working.

## 4 The Two-Nucleon Potential in Momentum Space

We are now in position to calculate any process involving soft pions and non-relativistic nucleons. Equations (1), (2) and (3) guarantee that the dominant contributions to such processes come from tree graphs with the maximum number of connected pieces and constructed out of the lagrangian $\mathcal{L}^{(0)}$. When applied to processes with at most one nucleon, this is equivalent to that given by current algebra. For example, the Weinberg [22] pion-pion and Tomozawa-Weinberg [22, 23] pion-nucleon $s$-wave scattering lengths are readily obtained. But in the late 1970s Weinberg [24] pointed out that chiral lagrangians, in addition, provide a framework for evaluating corrections to the dominant contributions. The systematic treatment of chiral perturbation theory in the mesonic sector began with the work of Gasser and Leutwyler [25] and has been extensively studied in the case of $SU(3) \times SU(3)$, up to $L = 1$ and $\Delta_i = 2$, and including electroweak effects (for an introduction, see Ref. [26]). A systematic study of the $SU(2) \times SU(2)$ chiral lagrangian for processes involving one nucleon was started by Gasser, Sainio and Švarc [27] and is continuing with the work of Bernard, Kaiser and Meißner, and many others (for a review see Ref.[28]). In principle, the coefficients $g_A$, $h_A$, $B_i$ and $A'_i$ can be determined from analyses of one nucleon processes once all contributions through one loop are evaluated. Unfortunately, this has not yet been done. In Sec. 7 we obtain values for all the parameters by fitting low energy nucleon-nucleon data. It should be kept in mind, however, that the number of parameters in the present potential could be reduced when sufficient information from the one-nucleon sector is gathered. For the many nucleon system the present theory is consistent with the empirical observation that three-(and more-)body forces are smaller than two-body forces. Some of the implications of this result are discussed in Ref. [29]. Furthermore, meson exchange currents [30], pion scattering [31] and pion photoproduction [32] on nuclei have also been studied in the same approach. For the remainder of this work we restrict our study to the two-nucleon system.

For only two nucleons in the initial and final state $A = 2$ and $C = 1$; Eq. (1) then simplifies to
$$\nu = 2L + \sum_i V_i \Delta_i \ . \tag{14}$$

As usual we work in the center-of-mass (c.m.) system and denote the initial energy by $2m_N + E$, initial (final) momentum by $\vec{p}(\vec{p}\,')$ and define $\vec{q} \equiv \vec{p} - \vec{p}\,'$ and $\vec{k} \equiv$ $\frac{1}{2}(\vec{p}+\vec{p}\,')$ as the transferred and average momenta, respectively. Subscripts 1 and 2 on spin and isospin matrices $\vec{\sigma}$ and $\boldsymbol{t}$ refer to nucleons 1 and 2.

The leading order potential $V^{(0)}$ (with $\nu = 0$) is obtained from the graphs in Fig.1 and interactions given by $\mathcal{L}^{(0)}$ in Eq. (11). Note that to this order nucleons are static, so that their energies in intermediate states are simply $m_N$ and the $\Delta$ isobar does not contribute. One obtains [16] the well-known static one-pion-exchange (OPE) potential supplemented by contact interactions where,

$$V^{(0)} = -(\frac{2g_A}{F_\pi})^2 \boldsymbol{t}_1 \cdot \boldsymbol{t}_2 \frac{\vec{\sigma}_1 \cdot \vec{q}\, \vec{\sigma}_2 \cdot \vec{q}}{\vec{q}^{\,2} + m_\pi^2} + C_S + C_T \vec{\sigma}_1 \cdot \vec{\sigma}_2 \ . \tag{15}$$

The OPE term provides the longest range part of the NN force, and it is well established [33] that it accounts for the higher partial waves in nucleon-nucleon scattering and the bulk of the properties of the deuteron, such as its quadrupole moment. Of course the NN potential has other sizeable components, including a spin-orbit force, a strong short-range repulsion and an intermediate range attraction. Clearly, the lowest order result in Eq. (15) does not account for these additional components. A test of the present approach is to determine whether higher order contributions yield such features.

First-order corrections in $Q/M$ ($\nu = 1$) also come from the graphs of Fig.1, but with one vertex from $\mathcal{L}^{(1)}$. However, there are no suitable vertices in Eq. (12) for the tree graphs in Fig.1 and we conclude that there are no corrections to the leading order potential $V^{(0)}$ that are smaller by just one power of $Q/M$, i.e.

$$V^{(1)} = 0. \tag{16}$$

This is a direct consequence of parity invariance. For the tree graphs, we could only add a power of momentum (or subtract one and add an extra power of $m_\pi^2$) to $V^{(0)}$, but this is actually a three-momentum because we eliminated time derivatives. This results in an odd number of three-momenta from which parity conserving terms cannot be constructed.

There are, however, many corrections of second-order, where $\nu = 2$. This includes tree graph contributions from $\mathcal{L}^{(2)}$ and a number of one-loop diagrams.

First, we obtain corrections from the tree graphs in Fig.1 where one vertex comes from the interactions in $\mathcal{L}^{(2)}$ in Eq. (13) and the nucleons remain static. We also obtain tree level corrections where the vertices are from $\mathcal{L}^{(0)}$ in Eq. (11) but where recoil is included in the intermediate state. Order $\nu = 2$ tree level corrections using two factors from $\mathcal{L}^{(1)}$ cannot be formed because, as we noticed above, there are no suitable vertices in Eq. (12). The tree graph $O[(Q/M)^2]$ correction is therefore given by

$$V_{tree}^{(2)} = -\frac{2g_A}{F_\pi^2} \boldsymbol{t}_1 \cdot \boldsymbol{t}_2 \frac{\vec{\sigma}_1 \cdot \vec{q}\, \vec{\sigma}_2 \cdot \vec{q}}{\vec{q}^{\,2} + m_\pi^2}$$
$$\times (A_1 q^2 + A_2 k^2 - 2g_A \frac{E - \frac{1}{4m_N}(4\vec{k}^2 + \vec{q}^{\,2})}{\sqrt{\vec{q}^{\,2} + m_\pi^2}})$$

$$+C_1\vec{q}^{\,2} + C_2\vec{k}^2 + (C_3\vec{q}^{\,2} + C_4\vec{k}^2)\vec{\sigma}_1 \cdot \vec{\sigma}_2$$
$$+iC_5\frac{\vec{\sigma}_1 + \vec{\sigma}_2}{2} \cdot (\vec{q} \times \vec{k}) + C_6\vec{q}\cdot\vec{\sigma}_1\vec{q}\cdot\vec{\sigma}_2$$
$$+C_7\vec{k}\cdot\vec{\sigma}_1\vec{k}\cdot\vec{\sigma}_2 \quad , \tag{17}$$

where the $A_i$'s and $C_i$'s are combinations (see Appendix B) of the $A_i'$'s and $C_i'$'s of Eq. (13). The explicit energy dependent term is discussed in Appendix C.

Second, there are contributions from the one-loop graphs in Fig.2 with all vertex factors coming from $\mathcal{L}^{(0)}$. (Other one loop graphs only contribute to the renormalization of parameters in the lagrangian.) Intermediate states include those with two nucleons, one nucleon and one isobar, and two isobars. Denoting

$$\omega_\pm \equiv \sqrt{(\vec{q}\pm\vec{l})^2 + 4m_\pi^2} \tag{18}$$
$$\Delta \equiv m_\Delta - m_N \quad , \tag{19}$$

straightforward calculation gives

$$\begin{aligned}
V^{(2)}_{loop,no\Delta} &= -\frac{1}{2F_\pi^4}\boldsymbol{t}_1\cdot\boldsymbol{t}_2 \int \frac{d^3l}{(2\pi)^3}\frac{1}{\omega_+\omega_-}\frac{(\omega_+-\omega_-)^2}{(\omega_++\omega_-)} \\
&\quad -4\left(\frac{g_A}{F_\pi^2}\right)^2\boldsymbol{t}_1\cdot\boldsymbol{t}_2 \int \frac{d^3l}{(2\pi)^3}\frac{1}{\omega_+\omega_-}\left(\frac{\vec{q}^{\,2}-\vec{l}^{\,2}}{\omega_+-\omega_-}\right) \\
&\quad -\frac{1}{4}\left(\frac{g_A}{F_\pi}\right)^4\int\frac{d^3l}{(2\pi)^3}\frac{1}{\omega_+^3\omega_-}\left\{\left(\frac{3}{\omega_-}+\frac{8\boldsymbol{t}_1\cdot\boldsymbol{t}_2}{\omega_++\omega_-}\right)(\vec{q}^{\,2}-\vec{l}^{\,2})^2\right. \\
&\quad \left. + 4\left(\frac{3}{\omega_++\omega_-}+\frac{8\boldsymbol{t}_1\cdot\boldsymbol{t}_2}{\omega_-}\right)\vec{\sigma}_1\cdot(\vec{q}\times\vec{l})\vec{\sigma}_2\cdot(\vec{q}\times\vec{l})\right\} \tag{20}
\end{aligned}$$

for the diagrams of Fig.2a,b,c,d that do not include isobars in intermediate states,

$$\begin{aligned}
V^{(2)}_{loop,one\Delta} &= \frac{8}{9}\frac{h_A^2}{F_\pi^4}\boldsymbol{t}_1\cdot\boldsymbol{t}_2\int\frac{d^3l}{(2\pi)^3}\frac{1}{(\omega_++\omega_-)}\frac{\vec{q}^{\,2}-\vec{l}^{\,2}}{(\omega_++2\Delta)(\omega_-+2\Delta)} \\
&\quad -\frac{1}{18}\frac{g_A^2 h_A^2}{F_\pi^4}\left\{(3+4\boldsymbol{t}_1\cdot\boldsymbol{t}_2)\right. \\
&\qquad \times \int\frac{d^3l}{(2\pi)^3}[(\vec{q}^{\,2}-\vec{l}^{\,2})^2 + 2\vec{\sigma}_1\cdot(\vec{q}\times\vec{l})\vec{\sigma}_2\cdot(\vec{q}\times\vec{l})] \\
&\qquad \times \left[\frac{1}{\omega_+\omega_-(\omega_++\omega_-)}\left(\frac{1}{\omega_+(\omega_-+2\Delta)}+\frac{1}{\omega_-(\omega_++2\Delta)}\right)\right. \\
&\qquad +\frac{1}{2}\frac{1}{\Delta\omega_+\omega_-}\left(\frac{1}{\omega_+\omega_-}+\frac{1}{\omega_+(\omega_-+2\Delta)}+\frac{1}{\omega_-(\omega_++2\Delta)}\right. \\
&\qquad \left.\left. + \frac{1}{(\omega_++2\Delta)(\omega_-+2\Delta)}\right)\right] \\
&\quad +(3-4\boldsymbol{t}_1\cdot\boldsymbol{t}_2)\int\frac{d^3\vec{l}}{(2\pi)^3}[(\vec{q}^{\,2}-\vec{l}^{\,2})^2 - 2\vec{\sigma}_1(\vec{q}\times\vec{l})\vec{\sigma}_2\cdot(\vec{q}\times\vec{l})]
\end{aligned}$$

$$\times \frac{1}{\omega_+\omega_-} \left[ \frac{1}{\omega_+ + \omega_- + 2\Delta} \left( \frac{1}{\omega_+\omega_-} + \frac{1}{(\omega_+ + 2\Delta)(\omega_- + 2\Delta)} \right) \right.$$

$$+ \left( \frac{1}{\omega_+ + \omega_-} + \frac{1}{\omega_+ + \omega_- + 2\Delta} \right)$$

$$\left. \times \left( \frac{1}{\omega_-(\omega_- + 2\Delta)} + \frac{1}{\omega_+(\omega_- + 2\Delta)} \right) \right] \Bigg\} \quad (21)$$

for the diagrams of Fig.2b,c,d,e with one intermediate isobar, and

$$V_{loop,two\Delta}^{(2)} = -\frac{2h_A^4}{81F_\pi^4} \Big\{ (3 - 2\boldsymbol{t}_1 \cdot \boldsymbol{t}_2)$$

$$\times \int \frac{d^3l}{(2\pi)^3} [(\vec{q}^2 - \vec{l}^2)^2 - \vec{\sigma}_1 \cdot (\vec{q} \times \vec{l}) \vec{\sigma}_2 \cdot (\vec{q} \times \vec{l})]$$

$$\times \frac{1}{\omega_+\omega_-} \frac{1}{(\omega_+ + 2\Delta)} \frac{1}{(\omega_- + 2\Delta)} \left[ \frac{1}{\omega_+ + \omega_-} + \frac{1}{2\Delta} \right]$$

$$+ (3 + 2\boldsymbol{t}_1 \cdot \boldsymbol{t}_2) \int \frac{d^3l}{(2\pi)^3} [(\vec{q}^2 - \vec{l}^2)^2 + \vec{\sigma}_1 \cdot (\vec{q} \times \vec{l}) \vec{\sigma}_2 \cdot (\vec{q} \times \vec{l})]$$

$$\times \frac{1}{\omega_+\omega_-(\omega_+ + \omega_- + 4\Delta)} \left[ \frac{1}{(\omega_+ + 2\Delta)(\omega_- + 2\Delta)} \right.$$

$$\left. + \frac{\omega_+ + \omega_- + 2\Delta}{\omega_+ + \omega_-} \left( \frac{1}{(\omega_- + 2\Delta)^2} + \frac{1}{(\omega_+ + 2\Delta)^2} \right) \right] \Bigg\} \quad (22)$$

for the diagrams of Fig. 2c,d,e that have two intermediate $\Delta$'s.

Finally, we consider corrections of order $[(Q/M)^3]$ where $\nu = 3$. Again, some terms could come from the tree graphs of Fig.1 with one vertex from $\mathcal{L}^{(3)}$, but the same argument used for $V^{(1)}$ guarantees that

$$V_{tree}^{(3)} = 0 \ . \quad (23)$$

Other third-order corrections would come from the one-loop graphs of Fig.2 where one vertex is from $\mathcal{L}^{(1)}$ in Eq. (12). Parity invariance requires the contribution from Fig.2a to vanish, as can be confirmed by explicit calculation, and because there are no $\pi \bar{N} N$ couplings in $\mathcal{L}^{(1)}$, the diagrams in Fig.2c,d,e also do not contribute. Fig.2b gives

$$V_{loop,no\Delta}^{(3)} = -\frac{1}{4} \left( \frac{g_A}{F_\pi^2} \right)^2 \int \frac{d^3l}{(2\pi)^3} \frac{1}{\omega_+^2 \omega_-^2} \Big\{ 3(\vec{q}^2 - \vec{l}^2)[-B_1(\vec{q}^2 - \vec{l}^2) + 4m_\pi^2 B_3]$$

$$+ 16 B_2 \vec{\sigma}_1 \cdot (\vec{q} \times \vec{l}) \vec{\sigma}_2 \cdot (\vec{q} \times \vec{l}) \boldsymbol{t}_1 \cdot \boldsymbol{t}_2 \Big\} \quad (24)$$

for no $\Delta$ in the intermediate state, and

$$V_{loop,one\Delta}^{(3)} = -\frac{1}{9} \left( \frac{h_A}{F_\pi^2} \right)^2 \int \frac{d^3l}{(2\pi)^3} \frac{1}{\omega_+\omega_-} \frac{1}{(\omega_+ + \omega_-)} \frac{1}{(\omega_+ + 2\Delta)(\omega_- + 2\Delta)}$$

$$\times \left\{ (\omega_+ + \omega_- + 2\Delta)[3(\vec{q}^{\,2} - \vec{l}^{\,2})(-B_1(\vec{q}^{\,2} - \vec{l}^{\,2}) + 4m_\pi^2 B_3) \right.$$
$$+ 4B_2 \vec{\sigma}_1 \cdot (\vec{q} \times \vec{l}) \vec{\sigma}_2 \cdot (\vec{q} \times \vec{l}) \boldsymbol{t}_1 \cdot \boldsymbol{t}_2]$$
$$\left. + 6B_1 \Delta \omega_+ \omega_- (\vec{q}^{\,2} - \vec{l}^{\,2}) \right\} \quad (25)$$

when there is one.

Further corrections are of higher order ($\nu \geq 4$). They include i) two-loop graphs, like the ones in Fig.3, that are numerous and harder to calculate, and ii) tree graphs with a vertex from $\mathcal{L}^{(4)}$, which would bring *many* new undetermined coefficients. We do not attempt to include them here.

The momentum space form of the potential, first presented in [14], facilitates a discussion of its structure and the comparison with other models. As usual the longest range part of the potential is given by one pion exchange [Eq. (15)], including the dominant, static OPE potential first obtained by Yukawa [1], plus corrections [Eq. (17)]. The $A_1$ and $A_2$ terms in Eq. (17) derive from the leading corrections to the $\pi \bar{N} N$ vertex that arise in an expansion of its form factor in powers of momenta over the form factor parameter. The $q^2$ dependence is usual (see for example Ref.[8] where monopole and dipole forms are used), whereas the $k^2$ dependence is not as common, however it too has been recently considered (*e.g.* Williamsburg model [34]). The other correction to the static OPE potential is the energy dependent term in Eq. (17), which arises from the recoil of the nucleon upon pion emission.

The intermediate range parts of the potential are due to two pion exchange (TPE) and are determined by parameters $F_\pi$, $g_A$, $h_A$, $m_\Delta - m_N$, $B_1$, $B_2$ and $B_3$. The contributions from box and crossed box diagrams (Fig.2c,d,e) are standard. The one in Eq. (20) ($g_A^4$ term) was first considered by Brueckner and Watson [2], while those with $\Delta$'s in Eq. (21) ($g_A^2 h_A^2$ terms) and Eq. (22) ($h_A^4$ terms) are due to Sugawara and von Hippel [4]. As a check, our results also agree with the appropriate limit of the expressions listed in Ref. [35]. But we would like to emphasize that there also exist TPE contributions from the "pair" diagrams of Fig.2a,b that are less common. Those in Eq. (20) and the $B_3$–term in Eq. (24) have also been suggested before by Sugawara and Okubo [3], but with arbitrary coefficients. Here the terms in Eq. (20) are fixed by chiral symmetry in terms of $g_A$ and $F_\pi$ while the $B_3$ term comes from the $\pi N$ $\sigma$-term. To the same order, we also have in Eq. (24) two new terms ($B_1, B_2$). The corresponding terms with $\Delta$ in Eqs. (21) and (25) are also new. It is important to emphasize that these contributions from the non-linear coupling of the pion to the nucleon are a consequence of chiral symmetry and that they are *not* usually included [2] in meson exchange potentials (*e.g.* Refs. [7, 8]). On the other hand, these terms are the only form of "correlated" pion exchange in our potential. The more traditional $s$-wave correlated TPE (Fig.3a) is higher order in the formalism discussed here.

---

[2] More recently, there has been some interest in the constraints of chiral symmetry to the TPE NN force, but limited to the diagrams corresponding to Eq. (20). For example, in Ref. [36] the scalar- isoscalar component of these diagrams has been studied, although a different definition of potential is considered; Ref. [37] discussed the relevance of energy dependence in OPEP to the definition of these TPE potentials; and in Ref. [38], Eq. (20) was examined for the unphysical case of $g_A = 1$.

The loop integrals in Eqs. (20), (21), (22), (24) and (25) diverge. Moreover, iteration in the Lippmann-Schwinger equation of (even the lowest order terms in) the potential produces further infinities. Regularization is therefore necessary, and counterterms are required to absorb the dependence on the regulator. The contact terms [the $C_i$'s in Eqs. (15) and (17)] perform exactly this function. Once renormalized, they contain the effect of exchange of higher energy modes and are not constrained by chiral symmetry; *i.e.* all combinations of spin operators and momenta (up to second power) that satisfy parity and time-reversal are included. This results in spin-orbit ($C_5$), spin-spin and tensor ($C_T, C_3, C_4, C_6, C_7$), and spin independent central ($C_S, C_1, C_2$) forces. In order to compare with other approaches it will be convenient to "undo" our previous Fierz reordering [21] and rewrite the coefficients $C_i$ as

$$C_i = C_i^{(0)} + C_i^{(1)} \bm{t}_1 \cdot \bm{t}_2. \tag{26}$$

# 5 The Two-Nucleon Potential in Coordinate - Space

Nucleon-nucleon scattering calculations, including those presented here, very often use a coordinate space representation. In order to transform the momentum space potential in Eqs. (15) – (26) into coordinate space we first have to specify the regularization procedure. The use of dimensional regularization here poses a problem that we have not yet succeeded in solving: how to iterate the potential to all orders in arbitrary dimension. Instead we use a momentum space cut-off $\Lambda \lesssim M$, as has been done in other potential models, because it is conceptually and mathematically simpler. The form of the cut-off function and the value assumed for $\Lambda$ are somewhat arbitrary and presumably not very important (see results in Sec. 7); variations in the cut-off are compensated to some extent by a redefinition of the free parameters in the theory. Again for simplicity, we follow the Nijmegen group [7] and assume a gaussian cut-off function $\exp(-\vec{l}^2/\Lambda^2)$, which regulates the loop integrals in the potential. In order to further regulate the loops arising from the iteration of the potential, we also cut-off the transferred momentum $q$ using the same cut-off function, $\exp(-\vec{q}^2/\Lambda^2)$.

All integrals over $\vec{q}$ and $\vec{l}$ can be reduced to simpler expressions involving one dimensional integrals that can easily be evaluated numerically. We use the formulas and techniques presented in Refs. [35, 39]—see Appendix D for details. Only the final form is presented here.

The tensor, total spin, and relative orbital angular momentum operators are defined, as usual, by

$$S_{12} = 3\frac{\vec{\sigma}_1 \cdot \vec{r} \vec{\sigma}_2 \cdot \vec{r}}{r^2} - \vec{\sigma}_1 \cdot \vec{\sigma}_2 \ ,$$

$$\vec{S} = \frac{1}{2}(\vec{\sigma}_1 + \vec{\sigma}_2) ,$$
$$\vec{L} = -i\vec{r} \times \vec{\nabla} , \quad (27)$$

respectively. In terms of these operators and the Pauli matrices $\boldsymbol{\tau}$ in isospin space the present potential can be expressed in terms of the following 20 operators:

$$\begin{aligned}\mathcal{O}^{p=1,\ldots,20} =\ & 1, \boldsymbol{\tau}_1\cdot\boldsymbol{\tau}_2, \vec{\sigma}_1\cdot\vec{\sigma}_2, \vec{\sigma}_1\cdot\vec{\sigma}_2\boldsymbol{\tau}_1\cdot\boldsymbol{\tau}_2, S_{12}, S_{12}\boldsymbol{\tau}_1\cdot\boldsymbol{\tau}_2, \vec{L}\cdot\vec{S}, \\ & \vec{L}\cdot\vec{S}\boldsymbol{\tau}_1\cdot\boldsymbol{\tau}_2, \vec{L}^2, \vec{L}^2\boldsymbol{\tau}_1\cdot\boldsymbol{\tau}_2, \vec{L}^2\vec{\sigma}_1\cdot\vec{\sigma}_2, \vec{L}^2\vec{\sigma}_1\cdot\vec{\sigma}_2\boldsymbol{\tau}_1\cdot\boldsymbol{\tau}_2, \\ & (\vec{L}\cdot\vec{S})^2, (\vec{L}\cdot\vec{S})^2\boldsymbol{\tau}_1\cdot\boldsymbol{\tau}_2, S_{12}\vec{L}\cdot\vec{S}, S_{12}\vec{L}\cdot\vec{S}\boldsymbol{\tau}_1\cdot\boldsymbol{\tau}_2, S_{12}\vec{L}^2, \\ & S_{12}\vec{L}^2\boldsymbol{\tau}_1\cdot\boldsymbol{\tau}_2, S_{12}(\vec{L}\cdot\vec{S})^2, S_{12}(\vec{L}\cdot\vec{S})^2\boldsymbol{\tau}_1\cdot\boldsymbol{\tau}_2 .\end{aligned} \quad (28)$$

The NN potential in coordinate space is written as

$$V = \sum_{p=1}^{20} V_p(r, \frac{\partial}{\partial r}, \frac{\partial^2}{\partial r^2}; E) \mathcal{O}^p \quad (29)$$

where

$$V_p(r, \frac{\partial}{\partial r}, \frac{\partial^2}{\partial r^2}; E) = V_p^0(r; E) + V_p^1(r; E)\frac{\partial}{\partial r} + V_p^2(r; E)\frac{\partial^2}{\partial r^2} \quad (30)$$

is an energy dependent radial operator determined by the radial functions $V_p^0(r;E)$, $V_p^1(r;E)$ and $V_p^2(r;E)$. These sixty functions (some vanish) are listed in Appendix E. Each consists of a sum of terms with coefficients determined by the parameters of the chiral lagrangian, and each term involves at most one one-dimensional integral of the functions from Appendix D. They are smooth at the origin thanks to regularization. The energy dependence in the radial functions of Eq. (30) is linear (see Appendix C).

The first eight operators, $\mathcal{O}^{p=1,\ldots,8}$, are standard and are accompanied in most potentials by radial functions with no derivatives. In this model they receive contributions from pion exchanges and contact terms. The next six operators, $\mathcal{O}^{p=9,\ldots,14}$, complete the set used in the phenomenological Urbana v14 potential [6], where, $V_p^1 = V_p^2 = 0$ for $p = 9, \ldots, 14$. What is characteristic of the structure of our potential is the presence of first and second derivative terms for $p = 1, \ldots 8$ and the presence of the other six operators $\mathcal{O}^{p=15,\ldots,20}$. All of these additional terms arise from the $O(k^2)$ dependence in the $A_2, C_2, C_4, C_7,$ and recoil correction terms.

# 6  Solution of the Schrödinger Equation

Having obtained a coordinate space representation of the potential the next step is to solve the Schrödinger equation numerically. The procedure is standard, but care must be exercised with respect to the derivative terms.

As usual, basis functions of definite total isospin $I$, total orbital angular momentum $L$, total spin $S$, and total angular momentum $J$ (and its third component $m$) were used; the relative c.m. NN wave function was decomposed into a partial wave sum of products of radial and spin-angle functions. By projecting onto the spin-angle basis a set of radial Schrödinger equations results which can be written schematically as

$$\left[ X^{(2)} \frac{\partial^2}{\partial r^2} + X^{(1)} \frac{\partial}{\partial r} + X^{(0)} \right] R = 0 , \qquad (31)$$

where

$$\begin{aligned} X^{(0)} &= \frac{1}{2\mu r^2} L(L+1) + \sum_p V_p^{(0)} \langle \mathcal{O}^p \rangle - E , \\ X^{(1)} &= -\frac{1}{\mu r} + \sum_p V_p^{(1)} \langle \mathcal{O}^p \rangle , \\ X^{(2)} &= -\frac{1}{2\mu} + \sum_p V_p^{(2)} \langle \mathcal{O}^p \rangle , \end{aligned} \qquad (32)$$

$\mu$ is the reduced mass, and $\langle \rangle$ denotes a matrix element between spin-angle basis functions. Spin singlet and triplet $L = J$ channels are uncoupled, so for these states $R$ is a single radial function. For the tensor coupled triplet states with $L = J \pm 1$, $R$ has two components and quantities $X^{(0)}$, $X^{(1)}$ and $X^{(2)}$ become $2 \times 2$ matrices.

In order to eliminate first derivative terms we define $R \equiv K\phi$ where the auxiliary function $K$ is chosen such that $\phi$ satisfies an equation with no first derivatives. This determines a differential equation for $K$ which depends on $X^{(1)}$ and $X^{(2)}$, given by

$$\frac{\partial K}{\partial r} = -\frac{1}{2} \left[ X^{(2)} \right]^{-1} X^{(1)} K , \qquad (33)$$

where $\mathrm{Det}(X^{(2)}) \neq 0$ and asymptotically $K \sim r^{-1}$. The boundary condition on $K$ for triplet channels is fixed by further requiring that the two components of $\phi$ are linearly independent as $r \to \infty$ which results in $\lim_{r \to \infty} K_{ij}(r) = \frac{1}{r}\delta_{ij}$, where $\delta_{ij}$ is the Kronecker delta function. Function $K$ at finite $r$ was obtained by Runge-Kutta integration of Eq. (33).

The resulting differential equation for $\phi$ is of the form

$$\frac{\partial^2 \phi(r)}{\partial r^2} = A(r; E)\phi(r) \qquad (34)$$

where $A(r; E)$ depends on $X^{(0)}$, $X^{(1)}$, $X^{(2)}$ and $K$. The wave function $\phi(r)$ satisfies the usual boundary conditions; i.e. $\phi$ vanishes at $r = 0$ and for large $r$, $\phi(r)$ matches to the asymptotic wave functions appropriate for scattering or bound states. The $S$-matrix or binding energy is obtained from the latter boundary condition. The NN $S$-matrix is expressed in terms of the usual phase shifts and mixing angles as in Eq. (7) of Ref. [40]. Eq. (34) was solved numerically for several positive scattering energies and for negative values of $E$ to determine the deuteron binding energy and other properties. The calculated phase shifts and deuteron properties depend on the undetermined parameters in the lagrangian. The cut-off parameter Λ was fixed and the remaining parameters of the lagrangian were varied until an optimized fit was obtained to recent NN phase shifts [41] (with errors from Ref. [42]) and measured deuteron properties [43].

# 7 Fitting Results for Phase Shifts and Deuteron Properties

The 26 parameters of the model ($g_A$, $h_A$, $F_\pi$, $A_1$, $A_2$, $B_1$, $B_2$, $B_3$, $C_S^{(0)}$, $C_T^{(0)}$, $C_1^{(0)}, \ldots C_7^{(0)}$, $C_S^{(1)}$, $C_T^{(1)}$, $C_1^{(1)}, \ldots C_7^{(1)}$, see Appendix E) were varied in order to optimize the fit to the isospin 0 (from $np$) and isospin 1 ($pp$) phase shifts of Ref. [41] at 10, 25, 50 and 100 MeV laboratory kinetic energy. All partial wave channels with total angular momentum $J \leq 2$ were included in the fits. In addition the $I = 0$, $^3S_1 - ^3D_1$ tensor coupled bound state (deuteron) binding energy, magnetic moment and electric quadrupole moment were also used to constrain the fit. The phase shifts for the $J > 2$ partial waves are dominated by the OPE potential at these low energies and were not used in the fitting procedure. The masses for the pion, nucleon and isobar used were $m_\pi = 140$ MeV, $m_N = 939$ MeV and $m_\Delta = 1232$ MeV, respectively. The principle results of this study were obtained assuming the cut-off parameter Λ to be 3.90 fm$^{-1}$ (equal to the $\rho$ mass). Sensitivity to the cut-off parameter is discussed later in this section.

The recent Nijmegen [41] phase shift solution was selected for fitting; errors were taken from the 1994 Arndt *et al.* [42] energy dependent phase shift analysis (solutions C10, C25, C50 and C100). The relative weighting of the chi-square contributions from the deuteron properties (binding energy, magnetic moment and electric quadrupole moment) and the scattering phase shifts was adjusted so as to achieve a suitable balance. The model was fitted to the phase shift parameters rather than directly to the NN scattering data since our goal here is to demonstrate the capabilities of the effective chiral lagrangian approach rather than to attempt to generate a phenomenological description of data which competes with other meson exchange models [7, 8, 9].

A grid search using parameters $h_A$, $A_1$, $A_2$, $B_1$, $B_2$, $B_3$, $C_S^{(0)}$, $C_T^{(0)}$, $C_1^{(0)}, \ldots C_7^{(0)}$ and fitting the $I = 0$ phase shifts and deuteron properties was initially conducted followed by a similar grid search for parameters $C_S^{(1)}$, $C_T^{(1)}$, $C_1^{(1)}, \ldots C_7^{(1)}$ for the $I = 1$ phase shifts using the previously optimized values of $h_A$, $A_1$, $A_2$, $B_1$, $B_2$, and $B_3$. The OPE $g_A$ and $F_\pi$ parameters were held fixed throughout the grid searches. A full, 26 parameter grid search was not feasible due to computational resource limitations. After locating a minimum in the chi-square space via the grid searches, the fits were optimized by simultaneously varying all 26 parameters using the downhill simplex method of chi-square minimization [44].

The best fits to the Nijmegen phase shifts with Λ = 3.90 fm$^{-1}$ are shown for $I = 0$ and 1 in Figs. 4 and 5, respectively. Except for a few of the channels at 100 MeV, the fits (solid lines) are in quantitative agreement with the phase shifts (data points) where the errors from [42] are shown if larger than the data symbol.

The results are essentially the same as shown previously in Ref. [15] but these new fits are in significantly better agreement with the 25 and 50 MeV $^1P_1$ and $\epsilon_1$ Nijmegen phases than was obtained in this earlier analysis of the older SP89 phase shift solution [42]. The $L = 0$ singlet and triplet scattering lengths are predicted by our model to be -15.6 and 5.40 fm, respectively, in comparison with the measured values of -16.4(1.9) fm [45] and 5.396(11) fm [46]. The optimized values obtained here for the 26 parameters are given in Table I.

The *predicted* phase shifts and mixing angles from our model (solid curves) for energies from 100 – 300 MeV are compared with the Nijmegen phase shift solutions (data points) in Figs. 6 and 7. For most of the partial wave parameters, except $^1P_1$, $\epsilon_1$ and $\epsilon_2$, the model predictions and phase shift solutions are in qualitative agreement. Because of the low momentum nature of the model, as expressed in the explicit $(Q/M)$ expansion, no effort was made to fit the phase shifts at energies above 100 MeV.

The deuteron properties for the $\Lambda = 3.90$ fm$^{-1}$ model fit are given in Table II in comparison with the measured values from Ref. [43]. Included are the binding energy, magnetic moment, electric quadrupole moment, asymptotic $d$-state to $s$-state wave function ratio, and $d$-state probability. The $s$- and $d$-state radial wave functions are also shown in Fig. 8. The negative portion of the $d$-state at small radii is also seen in the deuteron wave function of the Bonn potential [8], although both the radial extent and magnitude are larger here. We do not claim that the short range, high momentum components of our potential are realistic; no quantitative significance should be attached to this short-range part of the wave function. The depletion of the $d$-state at small radii, however, contributes to the low $d$-state probability of ∼3% which we obtain. Both the quadrupole moment and the asymptotic $d/s$ ratio are about 10% too small.

Also given in Table II are corrected values for the deuteron parameters corresponding to the potential model reported previously [15]. In these earlier calculations the deuteron wave function was computed incorrectly, resulting in erroneous values for the calculated magnetic moment, quadrupole moment and $d$-state probability [3]. For the corrected values the magnetic moment increased slightly by 1.4%, the quadrupole moment increased by 10% and is closer to the measured value, while the predicted $d$-state probability decreased from 5% to 3%. The scattering phase shifts, mixing angles, deuteron binding energy and asymptotic $d/s$ ratio in Ref. [15] are not affected.

It is interesting to study the sensitivity of the calculated phase shifts and deuteron parameters to the terms in the potential which are a direct consequence of chiral symmetry, corresponding to the diagrams in Figs. 2a and 2b. These include the first two terms in Eq. (20), the first term in Eq. (21) and the potentials in Eqs. (24) and (25) which depend on parameters $B_1$, $B_2$ and $B_3$. To study this sensitivity, calculations for all partial wave channels were made in which each of the above terms in the potential was individually set to zero. The first two terms in Eq. (20) and the first term in Eq. (21) have minor effects on the scattering phase shifts and mixing angles, however the chiral symmetry terms in Eq. (20) significantly affect the deuteron properties. The potentials in Eqs. (24) and (25), with the values for the parameters $B_1$, $B_2$ and $B_3$ given in Table I, contribute substantially to the scattering predictions and the deuteron. This applies to Eqs. (24) and (25) individually and to the $(B_1, B_2)$ terms and $B_3$ "$\sigma$-term" individually as well.

---

[3] We thank Prof. K. Holinde for suggesting there should be a mistake in our earlier value for the quadrupole moment.

The NN potential model presented here is, admittedly, complicated. To assist the reader we show in Fig. 9 the radial potentials for the $^1S_0$ channel corresponding to the $\Lambda = 3.90$ fm$^{-1}$ cut-off and the parameter values in Table I. The radial potentials $W^0$, $W^1$ and $W^2$ are defined by taking spin-angle matrix elements of the coordinate space potential in Eq. (29) where

$$\langle V \rangle \equiv W^0(r; E) + W^1(r; E) \frac{\partial}{\partial r} + W^2(r; E) \frac{\partial^2}{\partial r^2} \quad . \tag{35}$$

For coupled partial wave channels the $W$ functions become $2 \times 2$ matrices. The values for the $^1S_0$ potentials $W^0$, $W^1$ and $W^2$ at 50 MeV incident laboratory kinetic energy (the explicit energy dependence is weak) are shown in Fig. 9 by the dashed, dash-dot and dotted curves, respectively. The units for $W^0$, $W^1$ and $W^2$ are MeV, MeV·fm and MeV·fm$^2$, respectively. We also define an effective, local potential, $V_{eff}(r; E)$, according to:

$$A(r; E) \equiv \frac{L(L+1)}{r^2} + 2\mu V_{eff}(r; E) - 2\mu E \quad , \tag{36}$$

where $A(r; E)$ was defined in Eq. (34). The effective, local potential for this case is shown in Fig. 9 by the solid curve. If the first and second derivative terms in Eq. (35) were set to zero then $V_{eff}(r; E)$ would be identical to $W^0(r; E)$. The small difference between the solid and dashed curves in Fig. 9 is due to the derivative terms.

Fits to the phase shifts and deuteron properties were also obtained with cut-off parameter values of 2.50 fm$^{-1}$ and 5.00 fm$^{-1}$. The results for the phase shifts and mixing angles for the $\Lambda = 2.50$, 3.90 and 5.00 fm$^{-1}$ potentials are shown in Figs. 10 and 11 by the dashed, solid and dotted curves, respectively. Using the $\Lambda = 2.50$ fm$^{-1}$ cut-off the $^1P_1$ phase shift and $\epsilon_1$ mixing angle were better described than with the $\Lambda = 3.90$ fm$^{-1}$ cut-off, however poorer fits to the $^1S_0$, $^3P_2$ and $^3F_2$ phase shifts were obtained. Improved descriptions of the $^1P_1$ and $^3P_2$ phase shifts were achieved with the $\Lambda = 5.00$ fm$^{-1}$ cut-off value compared to the $\Lambda = 3.90$ fm$^{-1}$ results, however poorer descriptions of the $\epsilon_1$ and $\epsilon_2$ mixing angles and the $^1S_0$ and $^1D_2$ phase shifts resulted. The corresponding deuteron parameter values for $\Lambda = 2.50$ fm$^{-1}$ and 5.00 fm$^{-1}$ are also given in Table II. Overall we find qualitatively similar descriptions of the NN scattering results and deuteron properties for a wide range of cut-off parameters from 2.5 to 5.0 fm$^{-1}$ (corresponding to a mass range from 0.5 to 1.0 GeV).

# 8   Summary and Conclusions

We derived a low energy nucleon-nucleon potential, from an effective chiral lagrangian for soft pions and nonrelativistic nucleons using a perturbation expansion in powers of $(Q/M)$. We expressed the potential both in momentum space and in coordinate space, solved the corresponding Schrödinger equation in coordinate space, and fitted scattering phase shifts and deuteron properties by varying the undetermined parameters of the lagrangian.

In spirit, our approach is similar to that of the Paris group [9] where information on pion dynamics was used to construct the longer range parts of the potential, while more complicated dynamics was buried in unconstrained, short range parts. The fundamental difference between the approach of the Paris group and that of the present work is our use of effective field theory, rather than dispersion relations. Use of an effective chiral lagrangian not only ensures that our results are consistent with other aspects of pion phenomenology (chiral lagrangians to the order we use generally agree with data at the 20% level), but more importantly, explicitly incorporates the symmetries of QCD and provides a natural perturbative expansion. In this way we, like the Nijmegen group [7, 10], develop a potential within a theoretical framework, but unlike Refs. [7, 10] we carry out a controlled expansion. Our use of field theory and old-fashioned perturbation theory, on the other hand, causes our potential to be similar to a low-energy version of the Bonn potential [8].

The potential in momentum space shares several features with these and other potentials. The short range parts have all the necessary spin and isospin structure. The pion exchange terms result in contributions that have been considered before, but also result in several new terms related to chiral symmetry. Energy dependence (which has implications for few-body forces [29]) arises naturally.

The potential was transformed into coordinate space using a gaussian cut-off function. The $O(k^2)$ dependence in the momentum space potential leads to first and second derivative terms in the coordinate space representation. Elimination of first derivative terms in the radial Schrödinger equation through use of an auxiliary function permitted standard numerical methods to be employed.

We obtained reasonable, qualitative fits to the deuteron properties together with quantitative fits to most of the scattering phase shifts up to 100 MeV incident nucleon kinetic energy. This shows that our approach accounts for the principle features of the nucleon-nucleon potential and that these features can be naturally understood from the symmetries of QCD. However, the present work also makes clear that it is not practical for potential models derived from effective chiral lagrangians to compete with more phenomenological approaches, with respect to obtaining quantitative descriptions of NN data over a wide range of energies. Extension of the present model to higher energies and further improvement in the description of data could only result by including higher orders in chiral perturbation theory.

# Acknowledgements

We are grateful to many colleagues for discussions and comments, in particular R. A. Arndt, D. J. Ernst, K. Holinde, J. J. de Swart, R. Timmermans and S. Weinberg. This research was supported in part by U. S. Department of Energy Grants DE-FG03-94ER40845, DE-FG06-88ER40427 and DE-FG05-87ER40367 (with Vanderbilt University) and National Science Foundation grants NSF PHY 951 1632 and NSF PHY 900 1850 (with The University of Texas).

# A   Appendix

Pions are (pseudo)Goldstone bosons of the spontaneous breaking $SO(4) \to SO(3)$. They are associated with the broken generators of $SO(4)$ and therefore live in the sphere $SO(4)/SO(3) \sim S^3$. If we embed it in the euclidean $E^4$ space, $SO(4)$ transformations can be viewed as rotations of $S^3$ in $E^4$ planes. For example, $SU(2)_V$ of isospin consists of rotations in planes orthogonal to the fourth axis, while axial $SU(2)_A$ are rotations through planes that contain the fourth axis.

The sphere can be parametrized in a variety of ways, for example with four cartesian coordinates $\{\boldsymbol{\varphi}, \varphi_4 \equiv \sigma\}$ subject to the constraint,

$$\sigma^2 + \boldsymbol{\varphi}^2 = \frac{1}{4} F_\pi^2. \tag{37}$$

It is more convenient, however, to work with three unconstrained coordinates; therefore we use stereographic coordinates where

$$\boldsymbol{\pi} \equiv \frac{2\boldsymbol{\varphi}}{1 + \frac{2\sigma}{F_\pi}} . \tag{38}$$

Under an $SU(2)_V$ transformation with parameter $\boldsymbol{\varepsilon}$, the $\boldsymbol{\pi}$ coordinates rotate according to

$$\delta \boldsymbol{\pi} = \boldsymbol{\varepsilon} \times \boldsymbol{\pi}, \tag{39}$$

but they transform non-linearly under $SU(2)_A$ with parameter $\tilde{\boldsymbol{\varepsilon}}$ as given by

$$\delta \boldsymbol{\pi} = F_\pi \left(1 - \frac{\boldsymbol{\pi}^2}{F_\pi^2}\right) \frac{\tilde{\boldsymbol{\varepsilon}}}{2} + \frac{1}{F_\pi}(\tilde{\boldsymbol{\varepsilon}} \cdot \boldsymbol{\pi})\boldsymbol{\pi} . \tag{40}$$

A covariant derivative [see Eq. (4)] can be constructed, which is an isospin 1 object,

$$\delta \boldsymbol{D}_\mu = \boldsymbol{\varepsilon} \times \boldsymbol{D}_\mu, \tag{41}$$

which transforms under axial rotations as if under $SU(2)_V$ with a field-dependent parameter,

$$\delta \boldsymbol{D}_\mu = (\tilde{\boldsymbol{\varepsilon}} \times \frac{\boldsymbol{\pi}}{F_\pi}) \times \boldsymbol{D}_\mu. \tag{42}$$

Fermions also transform linearly under the unbroken subgroup

$$\delta N = i\boldsymbol{\varepsilon} \cdot \boldsymbol{t} N \tag{43}$$
$$\delta \Delta = i\boldsymbol{\varepsilon} \cdot \boldsymbol{t}^{(3/2)} \Delta. \tag{44}$$

In this case too, it is simplest to work with fields that realize the whole group non-linearly, i.e. that transform under axial transformations as if under isospin with the same field-dependent parameter as in Eq. (42). In this case

$$\delta N = i(\tilde{\boldsymbol{\varepsilon}} \times \frac{\boldsymbol{\pi}}{F_\pi}) \cdot \boldsymbol{t} N \tag{45}$$
$$\delta \Delta = i(\tilde{\boldsymbol{\varepsilon}} \times \frac{\boldsymbol{\pi}}{F_\pi}) \cdot \boldsymbol{t}^{(3/2)} \Delta . \tag{46}$$

It can be easily verified that the covariant derivatives of the pion, nucleon and isobar (Eqs. (7), (8) and (9), respectively) are indeed covariant; that is, they transform under $SU(2) \times SU(2)$ in the same way the fields $\boldsymbol{D}_\mu$, $N$ and $\Delta$ do (see Eqs. (41)—(46)).

A consequence of this is that an isoscalar constructed out of $\boldsymbol{D}_\mu$, $N$, $\Delta$ and their covariant derivatives will automatically be invariant under the whole $SU(2) \times SU(2)$. On the other hand, objects that transform under the full group as tensors involve also the $\boldsymbol{\pi}$ field itself. For example, an $SO(4)$ vector can be constructed according to

$$\left( \frac{2\frac{\boldsymbol{\pi}}{F_\pi}}{1 + \frac{\boldsymbol{\pi}^2}{F_\pi^2}}, \frac{1 - \frac{\boldsymbol{\pi}^2}{F_\pi^2}}{1 + \frac{\boldsymbol{\pi}^2}{F_\pi^2}} \right), \tag{47}$$

where its fourth component gives rise to the pion mass term in Eq. (11).

## B  Appendix

Here we list the relations between the $A_i$'s, $C_i$'s of Eq. (17) and the $A_i'$'s, and $C_i'$'s of Eq. (13):

$$\begin{aligned}
A_1 &= -(A_1' - \frac{1}{2}A_2') \\
A_2 &= -(A_1' + \frac{1}{2}A_2') \\
C_1 &= -C_1' + C_3' - \frac{1}{2}C_2' \\
C_2 &= 4(-C_1' + C_3' + \frac{1}{2}C_2') \\
C_3 &= -C_9' - \frac{1}{2}(C_{12}' + C_{14}') \\
C_4 &= 4(-C_9' + \frac{1}{2}(C_{12}' + C_{14}')) \\
C_5 &= -(2C_4' + C_5' - C_6') \\
C_6 &= -(C_7' + C_8' + \frac{1}{2}C_{10}' - C_{11}' - C_{13}') \\
C_7 &= -4(C_7' + C_8' - \frac{1}{2}C_{10}' + C_{11}' + C_{13}').
\end{aligned}$$

## C  Appendix

The origin of the explicit energy dependence of the present nucleon-nucleon potential is discussed here. One of the nice features of the chiral lagrangian approach is that it allows systematic inclusion of nucleon recoil corrections, *i.e.* energy dependent terms, as exemplified by Eq. (17) of Sec. 4. Here we give a somewhat general, though brief, description of how these terms arise. The systematic inclusion of recoil corrections has recently been shown to result in cancellations between reducible and irreducible graphs in the three-nucleon problem [29, 31]. This justifies, within this approach, certain approximations often made in nuclear physics.

The Lippmann-Schwinger equation for this case is given by

$$T_{AB}^{\tilde{E}\pm} = V_{AB} + \sum_C \frac{V_{AC} T_{CB}^{\tilde{E}\pm}}{\tilde{E}_B - \tilde{E}_C \pm i\epsilon} , \qquad (48)$$

where $\pm$ refer to outgoing and incoming wave boundary conditions,

$$V_{AB} = \left(\Phi_B, \hat{V}\Phi_A\right) , \qquad (49)$$

$$\hat{H} = \hat{H}_0 + \hat{V} , \qquad (50)$$

and the labels $A$, $B$ and $C$ denote quantum numbers for the free many-nucleon, pion, and isobar states $\Phi_A$. The energy parameters in Eq. (48) are the sum of the individual energies of these particles.

As is well known, Eq. (48) can be iterated to give the so-called "old-fashioned" perturbation theory, represented by the expansion

$$T_{AB}^{\tilde{E}} = V_{AB} + \sum_C V_{AC} \frac{1}{(\tilde{E}_B - \tilde{E}_C)} V_{CB} + \sum_{C,D} V_{AC} \frac{1}{(\tilde{E}_B - \tilde{E}_C)} V_{CD} \frac{1}{(\tilde{E}_B - \tilde{E}_D)} V_{DB} + \cdots , \qquad (51)$$

where the $\pm$ and the $i\epsilon$ are omitted to simplify the notation. Notice that $V_{AB}$ in Eq. (51) *is not* energy dependent.

Since we are interested in describing the low energy, nucleon-nucleon potential we choose the external particles to be only nonrelativistic nucleons. As in Sec. 2, it is convenient to introduce the effective potential as the sum of the irreducible diagrams of the series in Eq. (51). In the two-nucleon case this means diagrams where there is at least one pion or one isobar in the intermediate states (see Figs. 1 and 2). The complete set of diagrams can now be obtained by iterating this effective potential where the internal lines are two-nucleon lines ($A \to \alpha$, nucleons only):

$$T_{\alpha\beta}^{\tilde{E}} = V_{eff,\alpha\beta}(\tilde{E}) + \sum_\gamma V_{eff,\alpha\gamma}(\tilde{E}) \frac{1}{(\tilde{E}_\beta - \tilde{E}_\gamma)} V_{eff,\gamma\beta}(\tilde{E}) + \cdots \qquad (52)$$

Notice that $V_{eff,\alpha\beta}(\tilde{E})$ does depend, *by definition*, on the energy $\tilde{E}$ ($\tilde{E}_\beta$ in the energy denominators). To make contact with the nonrelativistic Schrödinger equation we recall that, for $n$ heavy nucleons,

$$\begin{aligned}
\tilde{E}_\beta - \tilde{E}_\alpha &= \sum_{i=1}^n \sqrt{m_N^2 + \mathbf{p}_i^2} - \sum_{i=1}^n \sqrt{m_N^2 + \mathbf{p}'^2_i} \\
&= \sum_{i=1}^n \frac{\mathbf{p}_i^2}{2m_N} - \sum_{i=1}^n \frac{\mathbf{p}'^2_i}{2m_N} + \text{small corrections} \\
&= E_\beta - E_\alpha + \text{small corrections} .
\end{aligned} \qquad (53)$$

Up to small corrections, which can be systematically accounted for, the effective potential depends on $E = \sum_i^n \mathbf{p}_i^2/(2m_N)$, the nonrelativistic kinetic energy. Clearly, in the infinite nucleon mass limit (static limit) this dependence vanishes and it is only in the $\mathcal{O}\left(\frac{Q}{M}\right)^2$ corrections to the lowest order term that they appear (Eq. (17)).

## D  Appendix

In order to obtain a potential in coordinate space we take Fourier transforms with a gaussian cut-off function with parameter $\Lambda$ (see [35, 39] for details). With

$$\mathrm{erfc}(x) = \frac{2}{\sqrt{\pi}} \int_x^\infty dt\, e^{-t^2}$$

denoting the complementary error function, we define and use the following functions:

$$I_0(r) = \frac{1}{8\pi\sqrt{\pi}} \Lambda^3 e^{-(\frac{r\Lambda}{2})^2}$$

$$I_2(r, m_\pi) = \frac{1}{8\pi r} e^{(\frac{m_\pi}{\Lambda})^2} \left[ e^{-m_\pi r} \mathrm{erfc}\left(-\frac{\Lambda r}{2} + \frac{m_\pi}{\Lambda}\right) - e^{m_\pi r} \mathrm{erfc}\left(\frac{\Lambda r}{2} + \frac{m_\pi}{\Lambda}\right) \right]$$

$$G_2(\lambda, r) = e^{-\frac{\lambda^2}{\Lambda^2}} I_2(r, \sqrt{m_\pi^2 + \lambda^2})$$

$$F_2(\lambda, r) = I_2(r, m_\pi) - G_2(\lambda, r)$$

$$\phi_C^0(r, m_\pi) = \frac{4\pi}{m_\pi} I_2(r, m_\pi)$$

$$\phi_C^1(r, m_\pi) = \phi_C^0(r, m_\pi) - \frac{4\pi}{m_\pi^3} I_0(r)$$

$$\phi_C^2(r, m_\pi) = \phi_C^1(r, m_\pi) + \frac{4\pi \Lambda^2}{m_\pi^5} \left[ \frac{3}{2} - \left(\frac{\Lambda r}{2}\right)^2 \right] I_0(r)$$

$$\phi_T^0(r, m_\pi) = \frac{1}{2(m_\pi r)^3} e^{(\frac{m_\pi}{\Lambda})^2} \left[ \left(1 + m_\pi r + \frac{1}{3}(m_\pi r)^2\right) e^{-m_\pi r} \mathrm{erfc}\left(-\frac{\Lambda r}{2} + \frac{m_\pi}{\Lambda}\right) \right.$$
$$\left. - \left(1 - m_\pi r + \frac{1}{3}(m_\pi r)^2\right) e^{m_\pi r} \mathrm{erfc}\left(\frac{\Lambda r}{2} + \frac{m_\pi}{\Lambda}\right) \right]$$
$$- \frac{4\pi}{3 m_\pi^3} \left(1 + \frac{6}{\Lambda^2 r^2}\right) I_0(r)$$

$$\phi_T^1(r, m_\pi) = \phi_T^0(r, m_\pi) - \frac{\pi r^2 \Lambda^4}{3 m_\pi^5} I_0(r)$$

$$\Sigma_1(r, \lambda) = m_\pi^3 \phi_T^0(r, m_\pi) - (m_\pi^2 + \lambda^2)^{3/2} e^{-\frac{\lambda^2}{\Lambda^2}} \phi_T^0(r, \sqrt{m_\pi^2 + \lambda^2})$$

$$\Sigma_2(r, \lambda) = \frac{1}{3} m_\pi^3 \phi_C^1(r, m_\pi) - \frac{1}{3}(m_\pi^2 + \lambda^2)^{3/2} e^{-\frac{\lambda^2}{\Lambda^2}} \phi_C^1(r, \sqrt{m_\pi^2 + \lambda^2})$$

$$\Omega_1(r, \lambda) = m_\pi^5 \phi_T^1(r, m_\pi) - (m_\pi^2 + \lambda^2)^{5/2} e^{-\frac{\lambda^2}{\Lambda^2}} \phi_T^1(r, \sqrt{m_\pi^2 + \lambda^2})$$

$$\Omega_2(r,\lambda) = \frac{1}{3}m_\pi^5 \phi_C^2(r, m_\pi) - \frac{1}{3}(m_\pi^2 + \lambda^2)^{5/2} e^{-\frac{\lambda^2}{\Delta^2}} \phi_C^2(r, \sqrt{m_\pi^2 + \lambda^2}) \ ;$$

plus the integrals:

$$\mathcal{R}_\Delta^{(n,m)}[f] = \frac{2}{\pi} \int_0^\infty d\lambda \frac{\lambda^{2m}}{(\lambda^2 + \Delta^2)^n} f(\lambda)$$

$$\mathcal{H}_1(r) = \Delta \mathcal{R}_\Delta^{(1,0)}[G_2]$$

$$\mathcal{H}_2(r) = \frac{1}{\Delta}[I_2(r, m_\pi) - \mathcal{H}_1(r)] \ ,$$

where $f$ is any function of $\lambda$ and $\Delta = m_\Delta - m_N$.

# E  Appendix

Here we give the explicit forms of the 60 radial potential functions $V_p^i(r)$, $p = 1, \ldots, 20$; $i = 0, 1, 2$, which appear in the coordinate space version of the potential in Eqs. (29) and (30). To save space the following combinations of functions and derivatives of functions are defined:

$$D_1(f) \equiv \frac{f'}{r}\left(2f'' + \frac{1}{r}f'\right)$$

$$D_2(f) \equiv \frac{f'}{r}\left(f'' - \frac{1}{r}f'\right)$$

$$\varepsilon_1(f) \equiv f' + \frac{2}{r}f$$

$$\varepsilon_2(f) \equiv f' - \frac{1}{r}f$$

$$\mathcal{S}(f,g) \equiv f''g'' + \frac{2}{r^2}f'g'$$

$$\mathcal{T}(f,g) \equiv \frac{1}{r}(f''g' + f'g'') + \frac{1}{r^2}f'g'$$

$$\mathcal{P}(f,g) \equiv \frac{2}{r^2}f'g' - \frac{1}{r}(f'g'' + f''g')$$

$$\mathcal{Q}(f,g) \equiv \left(-\frac{4}{r}f' - 2f'' + 2m_\pi^2 f - I_0\right) g \ ,$$

where $f = f(r)$ and $g = g(r)$ are any of the functions defined in Appendix D, and a prime denotes differentiation with respect to $r$.

We then have (where $\hbar c = 1$):

$$V_1^0 = \frac{1}{F_\pi^4}\Big\{-3g_A^4 \mathcal{R}_0^{(1,0)}[\mathcal{S}(I_2, F_2)] + 6g_A^2[B_1 \mathcal{S}(I_2, I_2) + B_3 m_\pi^2 (I_2')^2]$$

$$+\frac{2g_A^2 h_A^2}{3}\Big[\Delta \mathcal{S}(\mathcal{H}_2, \mathcal{H}_2) - 4\mathcal{S}(I_2, \mathcal{H}_2) - \frac{1}{\Delta}\mathcal{S}(I_2 + \mathcal{H}_1, I_2 + \mathcal{H}_1)\Big]$$

$$
\begin{aligned}
& -\frac{16}{27}h_A^4\Delta^2\mathcal{R}_\Delta^{(2,0)}[\mathcal{S}(G_2,G_2)] \\
& +\frac{8}{3}h_A^2\Delta\left[B_1\left(\mathcal{R}_\Delta^{(1,0)}[\mathcal{S}(G_2,G_2)]+\mathcal{R}_\Delta^{(1,1)}[G_2'^2]\right)+B_3m_\pi^2\mathcal{R}_\Delta^{(1,0)}[G_2'^2]\right]\bigg\} \\
& +C_S^{(0)}I_0-(C_1^{(0)}+\frac{1}{4}C_2^{(0)})\varepsilon_1(I_0')
\end{aligned}
$$

$$V_1^1 = -C_2^{(0)}\varepsilon_1(I_0)$$

$$V_1^2 = -C_2^{(0)}I_0$$

$$
\begin{aligned}
V_2^0 = & \frac{1}{F_\pi^4}\bigg\{-2g_A^4\mathcal{R}_0^{(1,0)}[\mathcal{S}(G_2,F_2)]-\mathcal{R}_0^{(0,0)}[\mathcal{Q}(G_2,G_2)]+4g_A^2\mathcal{R}_0^{(0,0)}[G_2'^2] \\
& +\frac{2}{9}g_A^2h_A^2\bigg[-3\Delta\mathcal{S}(\mathcal{H}_2,\mathcal{H}_2)+4\mathcal{S}(I_2,\mathcal{H}_2)+\frac{4}{\Delta}\mathcal{S}(\mathcal{H}_1,\mathcal{H}_1)-4\mathcal{R}_\Delta^{(1,0)}[\mathcal{S}(G_2,G_2)] \\
& -\frac{1}{\Delta}\mathcal{S}(I_2+\mathcal{H}_1,I_2+\mathcal{H}_1)\bigg]+\frac{8}{81}h_A^4\mathcal{R}_\Delta^{(2,1)}[\mathcal{S}(G_2,G_2)]-\frac{8}{9}h_A^2\mathcal{R}_\Delta^{(1,1)}[G_2'^2]\bigg\} \\
& +\frac{1}{4}C_S^{(1)}I_0-\frac{1}{4}(C_1^{(1)}+\frac{1}{4}C_2^{(1)})\varepsilon_1(I_0')
\end{aligned}
$$

$$V_2^1 = -\frac{1}{4}C_2^{(1)}\varepsilon_1(I_0)$$

$$V_2^2 = -\frac{1}{4}C_2^{(1)}I_0$$

$$
\begin{aligned}
V_3^0 = & \frac{1}{F_\pi^4}\bigg\{-2g_A^4\mathcal{R}_0^{(1,0)}[\mathcal{T}(G_2,F_2)] \\
& -\frac{2}{9}g_A^2h_A^2\bigg[3\Delta D_1(\mathcal{H}_2)-4\mathcal{T}(I_2,\mathcal{H}_2)-\frac{4}{\Delta}D_1(\mathcal{H}_1)+4\mathcal{R}_\Delta^{(1,0)}[D_1(G_2)] \\
& +\frac{1}{\Delta}D_1(I_2+\mathcal{H}_1)\bigg]+\frac{8}{81}h_A^4\mathcal{R}_\Delta^{(2,1)}[D_1(G_2)]\bigg\} \\
& +C_T^{(0)}I_0-\left(C_3^{(0)}+\frac{1}{4}C_4^{(0)}+\frac{1}{3}C_6^{(0)}+\frac{1}{12}C_7^{(0)}\right)\varepsilon_1(I_0')
\end{aligned}
$$

$$V_3^1 = -(C_4^{(0)}+\frac{1}{3}C_7^{(0)})\varepsilon_1(I_0)$$

$$V_3^2 = -(C_4^{(0)}+\frac{1}{3}C_7^{(0)})I_0$$

$$
\begin{aligned}
V_4^0 = & \frac{1}{F_\pi^4}\bigg\{-\frac{4}{3}g_A^4\mathcal{R}_0^{(1,0)}[\mathcal{T}(I_2,F_2)]-\frac{4}{3}g_A^2B_2D_1(I_2) \\
& +\frac{2}{27}g_A^2h_A^2\bigg[\Delta D_1(\mathcal{H}_2)-4\mathcal{T}(I_2,\mathcal{H}_2)-\frac{1}{\Delta}D_1(I_2+\mathcal{H}_1)\bigg] \\
& -\frac{4}{243}h_A^4\Delta^2\mathcal{R}_\Delta^{(2,0)}[D_1(G_2)]-\frac{4}{27}h_A^2B_2\Delta\mathcal{R}_\Delta^{(1,0)}[D_1(G_2)]\bigg\} \\
& +\left(\frac{2g_A}{F_\pi}\right)^2\bigg\{\frac{m_\pi^3}{48\pi}[\phi_C^1-\frac{A_1m_\pi^2}{2g_A}\phi_C^2]+\frac{1}{32}\left[E\mathcal{R}_0^{(1,0)}[\Sigma_2]+\frac{1}{4m_N}\mathcal{R}_0^{(1,0)}[\Omega_2]\right] \\
& +\frac{1}{128}\left[\frac{1}{m_N}\mathcal{R}_0^{(1,0)}[\varepsilon_1(\Sigma_2')]-\frac{A_2m_\pi^3}{3\pi g_A}\varepsilon_1(\phi_C^{1'})\right]\bigg\}
\end{aligned}
$$

$$
\begin{aligned}
&\quad +\frac{1}{4}C_T^{(1)}I_0 - \frac{1}{4}\left(C_3^{(1)} + \frac{1}{4}C_4^{(1)} + \frac{1}{3}C_6^{(1)} + \frac{1}{12}C_7^{(1)}\right)\varepsilon_1(I_0') \\
V_4^1 &= \frac{g_A^2}{8F_\pi^2}\left[\frac{1}{m_N}\mathcal{R}_0^{(1,0)}[\varepsilon_1(\Sigma_2)] - \frac{A_2 m_\pi^3}{3g_A\pi}\varepsilon_1(\phi_C^1)\right] - \frac{1}{4}(C_4^{(1)} + \frac{1}{3}C_7^{(1)})\varepsilon_1(I_0) \\
V_4^2 &= \frac{g_A^2}{8F_\pi^2}\left(\frac{1}{m_N}\mathcal{R}_0^{(1,0)}[\Sigma_2] - \frac{A_2 m_\pi^3}{3\pi g_A}\phi_C^1\right) - \frac{1}{4}(C_4^{(1)} + \frac{1}{3}C_7^{(1)})I_0 \\
V_5^0 &= \frac{1}{F_\pi^4}\Big\{-g_A^4 \mathcal{R}_0^{(1,0)}[\mathcal{P}(G_2, F_2)] + \frac{2}{9}g_A^2 h_A^2\Big[3\Delta D_2(\mathcal{H}_2) + 2\mathcal{P}(I_2, \mathcal{H}_2) - \frac{4}{\Delta}D_2(\mathcal{H}_1) \\
&\qquad\qquad + 4\mathcal{R}_\Delta^{(1,0)}[D_2(G_2)] + \frac{8}{\Delta}D_2(I_2 + \mathcal{H}_1)\Big] - \frac{8}{81}h_A^4 \mathcal{R}_\Delta^{(2,1)}[D_2(G_2)]\Big\} \\
&\quad - \frac{1}{3}(C_6^{(0)} + \frac{1}{4}C_7^{(0)})\varepsilon_2(I_0') \\
V_5^1 &= -\frac{1}{3}C_7^{(0)}\varepsilon_2(I_0) \\
V_5^2 &= -\frac{1}{3}C_7^{(0)}I_0 \\
V_6^0 &= \frac{1}{F_\pi^4}\Big\{-\frac{2}{3}g_A^4 \mathcal{R}_0^{(1,0)}[\mathcal{P}(I_2, F_2)] + \frac{4}{3}g_A^2 B_2 D_2(I_2) \\
&\qquad\qquad -\frac{2}{27}g_A^2 h_A^2\left(\Delta D_2(\mathcal{H}_2) + 2\mathcal{P}(I_2, \mathcal{H}_2) - \frac{1}{\Delta}D_2(I_2 + \mathcal{H}_1)\right) \\
&\qquad\qquad + \frac{4}{243}h_A^4 \Delta^2 \mathcal{R}_\Delta^{(2,0)}[D_2(G_2)] + \frac{4}{27}h_A^2 B_2 \Delta \mathcal{R}_\Delta^{(1,0)}[D_2(G_2)]\Big\} \\
&\quad + \frac{g_A^2}{F_\pi^2}\Big\{\frac{m_\pi^3}{4\pi}[\phi_T^0 - \frac{A_1 m_\pi^2}{2g_A}\phi_T^1] + \frac{1}{8}\left[E\mathcal{R}_0^{(1,0)}[\Sigma_1] + \frac{1}{4m_N}\mathcal{R}_0^{(1,0)}[\Omega_1]\right] \\
&\qquad + \frac{1}{32m_N}\mathcal{R}_0^{(1,0)}[\varepsilon_1(\Sigma_1') - \frac{6}{r^2}\Sigma_1] - \frac{A_2 m_\pi^3}{32\pi g_A}\left(\varepsilon_1(\phi_T^{0'}) - \frac{6}{r^2}\phi_T^0\right)\Big\} \\
&\quad - \frac{1}{12}(C_6^{(1)} + \frac{1}{4}C_7^{(1)})\varepsilon_2(I_0') \\
V_6^1 &= \frac{g_A^2}{8F_\pi^2}\Big\{\frac{1}{m_N}\mathcal{R}_0^{(1,0)}[\varepsilon_1(\Sigma_1) + \frac{2}{r}\Sigma_1] - \frac{A_2 m_\pi^3}{g_A \pi}\left(\varepsilon_1(\phi_T^0) + \frac{2}{r}\phi_T^0\right)\Big\} \\
&\quad - \frac{1}{12}C_7^{(1)}\varepsilon_2(I_0) \\
V_6^2 &= \frac{g_A^2}{4F_\pi^2}\left\{\frac{1}{m_N}\mathcal{R}_0^{(1,0)}[\Sigma_1] - \frac{A_2 m_\pi^3}{g_A \pi}\phi_T^0\right\} - \frac{1}{12}C_7^{(1)}I_0 \\
V_7^0 &= \frac{1}{r}\left[-C_5^{(0)}I_0' + \frac{1}{3}C_7^{(0)}\varepsilon_1(I_0)\right] \\
V_7^1 &= \frac{2}{3r}C_7^{(0)}I_0 \\
V_7^2 &= 0 \\
V_8^0 &= \left(\frac{g_A}{2rF_\pi}\right)^2\left[\frac{A_2 m_\pi^3}{\pi g_A}\phi_T^0 - \frac{1}{m_N}\mathcal{R}_0^{(1,0)}[\Sigma_1]\right] + \frac{1}{4r}\left[-C_5^{(1)}I_0' + \frac{1}{3}C_7^{(1)}\varepsilon_1(I_0)\right]
\end{aligned}
$$

$$
\begin{aligned}
V_8^1 &= \frac{1}{6r} C_7^{(1)} I_0 \\
V_8^2 &= 0 \\
V_9^0 &= C_2^{(0)} \frac{I_0}{r^2} \\
V_9^1 &= V_9^2 = 0 \\
V_{10}^0 &= \frac{1}{4} C_2^{(1)} \frac{I_0}{r^2} \\
V_{10}^1 &= V_{10}^2 = 0 \\
V_{11}^0 &= (C_4^{(0)} + \frac{1}{3} C_7^{(0)}) \frac{I_0}{r^2} \\
V_{11}^1 &= V_{11}^2 = 0 \\
V_{12}^0 &= \frac{1}{2} \left( \frac{g_A}{2r F_\pi} \right)^2 \left[ \frac{A_2 m_\pi^3}{3 g_A \pi} \phi_C^1 - \frac{1}{m_N} \mathcal{R}_0^{(1,0)}[\Sigma_2] \right] + \frac{1}{4} \left( C_4^{(1)} + \frac{1}{3} C_7^{(1)} \right) \frac{I_0}{r^2} \\
V_{12}^1 &= V_{12}^2 = 0 \\
V_{13}^0 &= -\frac{2}{3} C_7^{(0)} \frac{I_0}{r^2} \\
V_{13}^1 &= V_{13}^2 = 0 \\
V_{14}^0 &= -\frac{1}{6} C_7^{(1)} \frac{I_0}{r^2} \\
V_{14}^1 &= V_{14}^2 = 0 \\
V_{15}^0 &= \frac{1}{3} C_7^{(0)} \left( \frac{I_0'}{r} - 3 \frac{I_0}{r^2} \right) \\
V_{15}^1 &= \frac{2}{3} C_7^{(0)} \frac{I_0}{r} \\
V_{15}^2 &= 0 \\
V_{16}^0 &= \left( \frac{g_A}{2r F_\pi} \right)^2 \left[ \frac{A_2 m_\pi^3}{g_A \pi} \phi_T^0 - \frac{1}{m_N} \mathcal{R}_0^{(1,0)}[\Sigma_1] \right] + \frac{1}{12} C_7^{(1)} \left( \frac{I_0'}{r} - 3 \frac{I_0}{r^2} \right) \\
V_{16}^1 &= \frac{1}{6} C_7^{(1)} \frac{I_0}{r} \\
V_{16}^2 &= 0 \\
V_{17}^0 &= \frac{1}{3} C_7^{(0)} \frac{I_0}{r^2} \\
V_{17}^1 &= V_{17}^2 = 0 \\
V_{18}^0 &= \left( \frac{g_A}{2r F_\pi} \right)^2 \left[ \frac{A_2 m_\pi^3}{g_A \pi} \phi_T^0 - \frac{1}{m_N} \mathcal{R}_0^{(1,0)}[\Sigma_1] \right] + \frac{1}{12} C_7^{(1)} \frac{I_0}{r^2} \\
V_{18}^1 &= V_{18}^2 = 0 \\
V_{19}^0 &= -\frac{2}{3} C_7^{(0)} \frac{I_0}{r^2} \\
V_{19}^1 &= V_{19}^2 = 0
\end{aligned}
$$

$$\begin{aligned}
V_{20}^0 &= -\frac{1}{6} C_7^{(1)} \frac{I_0}{r^2} \\
V_{20}^1 &= V_{20}^2 = 0 \quad .
\end{aligned}$$

Table I: Effective chiral lagrangian potential model parameters for $\Lambda = 3.90$ fm$^{-1}$ based on the fit to the Nijmegen phase shifts [41].

| | | |
|---|---|---|
| $g_A$ | 1.33 | |
| $h_A$ | 2.03 | |
| $F_\pi$ (MeV) | 192 | |
| $A_1$ ($10^{-6}$ MeV$^{-2}$) | -1.38 | |
| $A_2$ ($10^{-6}$ MeV$^{-2}$) | 2.44 | |
| $B_1$ ($10^{-2}$ MeV$^{-1}$) | 0.342 | |
| $B_2$ ($10^{-2}$ MeV$^{-1}$) | 0.854 | |
| $B_3$ ($10^{-2}$ MeV$^{-1}$) | 1.77 | |
| | $I = 0$ | $I = 1$ |
| $C_S$ ($10^{-4}$ MeV$^{-2}$) | 1.12 | 0.135 |
| $C_T$ ($10^{-4}$ MeV$^{-2}$) | -0.266 | -0.689 |
| $C_1$ ($10^{-9}$ MeV$^{-4}$) | 0.661 | 0.381 |
| $C_2$ ($10^{-9}$ MeV$^{-4}$) | 3.39 | 2.97 |
| $C_3$ ($10^{-9}$ MeV$^{-4}$) | -0.330 | -0.0295 |
| $C_4$ ($10^{-9}$ MeV$^{-4}$) | -0.144 | 0.453 |
| $C_5$ ($10^{-9}$ MeV$^{-4}$) | 2.10 | -0.910 |
| $C_6$ ($10^{-9}$ MeV$^{-4}$) | 0.281 | 0.0998 |
| $C_7$ ($10^{-9}$ MeV$^{-4}$) | 0.581 | 1.36 |

Table II: Experimental and effective chiral lagrangian model fitted values for the deuteron binding energy ($BE$), magnetic moment ($\mu_d$), electric quadrupole moment ($Q_E$), asymptotic $d/s$ ratio ($\eta$), and $d$-state probability ($P_D$).

| Deuteron Quantities | Experiment[b] | Fit to Nijmegen phase shifts [41] | | | SP89 Fits[a] |
|---|---|---|---|---|---|
| | | $\Lambda = 2.50\text{fm}^{-1}$ | $\Lambda = 3.90\text{fm}^{-1}$ | $\Lambda = 5.00\text{fm}^{-1}$ | $\Lambda = 3.90\text{fm}^{-1}$ |
| $BE$ (MeV) | 2.224579(9) | 2.15 | 2.24 | 2.18 | 2.18 |
| $\mu_d$ ($\mu_N$) | 0.857406(1) | 0.863 | 0.863 | 0.866 | 0.863 |
| $Q_E$ (fm$^2$) | 0.2859(3) | 0.246 | 0.249 | 0.237 | 0.253 |
| $\eta$ | 0.0271(4) | 0.0229 | 0.0244 | 0.0230 | 0.0239 |
| $P_D$ (%) | | 2.98 | 2.86 | 2.40 | 2.89 |

[a]Corrected values given here for fit in Ref. [15]
[b]See Ref. [43]

# References

[1] H. Yukawa, Proc. Phys.-Math. Soc. Jpn. **17** (1935) 48.

[2] K.A. Brueckner and K.M. Watson, Phys. Rev. **92** (1953) 1023.

[3] M. Sugawara and S. Okubo, Phys. Rev. **117** (1960) 605; *ibid.* 611.

[4] H. Sugawara and F. von Hippel, Phys. Rev. **172** (1968) 1764.

[5] R.V. Reid, Ann. Phys. (N. Y.) **50** (1968) 411.

[6] I.E. Lagaris and V.R. Pandharipande, Nucl. Phys. **A359** (1981) 331.

[7] M.M. Nagels, T.A. Rijken and J.J. de Swart, Phys. Rev. D **17** (1978) 768; V.G.J. Stoks, R.A.M. Klomp, C.P.F. Terheggen and J.J. de Swart, Phys. Rev. C **49** (1994) 2950.

[8] R. Machleidt, K. Holinde and Ch. Elster, Phys. Rep. **149** (1987) 1; R. Machleidt, in *Advances in Nuclear Physics*, Vol. **19**, edited by J.W. Negele and E. Vogt (Plenum, New York, 1989), pp. 189-376.

[9] M. Lacombe, B. Loiseau, J.M. Richard, R. Vinh Mau, J. Côté, P. Pirès, and R. de Tourreil, Phys. Rev. C **21** (1980) 861; R. Vinh Mau, "The Paris Nucleon-Nucleon Potential," in *Mesons in Nuclei*, Vol. I, edited by M. Rho and D. Wilkinson, (North-Holland, Amsterdam, 1979), pp. 151-196.

[10] T.A. Rijken, Ann. Phys. **164** (1985) 1; *ibid.* 23.

[11] F. Myhrer and J. Wroldsen, Rev. Mod. Phys. **60** (1988) 629; K. Yazaki, "Baryon-Baryon Interactions — Quarks versus Mesons," in *Few-Body Problems in Physics*, edited by F. Gross, (AIP, New York, 1995), pp. 225-238.

[12] A. Jackson, A.D. Jackson and V. Pasquier, Nucl. Phys. **A432** (1985) 567; R. Vinh Mau, M. Lacombe, B. Loiseau, W.N. Cottingham and P. Lisboa, Phys. Lett. B **150** (1985) 259.

[13] J. Wambach, Acta Phys. Polon. B **23** (1992) 1163.

[14] C. Ordóñez and U. van Kolck, Phys. Lett. B **291** (1992) 459.

[15] C. Ordóñez, L. Ray and U. van Kolck, Phys. Rev. Lett. **72** (1994) 1982.

[16] S. Weinberg, Phys. Lett. B **251** (1990) 288.

[17] S. Weinberg, Nucl. Phys. **B363** (1991) 3.

[18] A. Manohar and H. Georgi, Nucl. Phys. **B234** (1984) 189.

[19] E. Jenkins and A.V. Manohar, Phys. Lett. B **255** (1991) 558.

[20] U. van Kolck, U. of Texas Ph.D. dissertation (1993), and U. of Washington preprint DOE/ER/40427-13-N94.

[21] M. Fierz, Z. Phys. **104** (1937) 553.

[22] S. Weinberg, Phys. Rev. Lett. **17** (1966) 616.

[23] Y. Tomozawa, Nuovo Cim. **46A** (1966) 707.

[24] S. Weinberg, Physica **96A** (1979) 327.

[25] J. Gasser and H. Leutwyler, Ann. Phys. **158** (1984) 142; Nucl. Phys. **B250** (1985) 465.

[26] G. Ecker, "Chiral Perturbation Theory," in *Quantitative Particle Physics, Cargèse 1992*, edited by M. Lévy, J.-L. Basdevant, M. Jacob, J. Iliopoulos, R. Gastmans and J.-M. Gérard, NATO ASI Series B, Vol. 311, (Plenum, New York, 1993), pp. 101-148.

[27] J. Gasser, M.E. Sainio and A. Švarc, Nucl. Phys. **B307** (1988) 779.

[28] V. Bernard, N. Kaiser and U.-G. Meißner, Int. J. Mod. Phys. **E4** (1995) 193.

[29] U. van Kolck, Phys. Rev. C **49** (1994) 2932.

[30] T.-S. Park, D.-P. Min and M. Rho, Phys. Rep. **233** (1993) 341.

[31] S. Weinberg, Phys. Lett. B **295** (1992) 114.

[32] S.R. Beane, C.Y. Lee and U. van Kolck, U. of Washington preprint DOE/ER/40427-06-N95, nucl-th/9506017.

[33] T.E.O. Ericson, Prog. Part. Nucl. Phys. **11** (1984) 245.

[34] F. Gross, J.W. van Orden and K. Holinde, Phys. Rev. C **45** (1992) 2094.

[35] Th.A. Rijken and V.G.J. Stoks, Phys. Rev. C **46** (1992) 73.

[36] L.S. Celenza, A. Pantziris and C.M. Shakin, Phys. Rev. C **46** (1992) 2213.

[37] J.L. Friar and S.A. Coon, Phys. Rev. C **49** (1994) 1272.

[38] C.A. da Rocha and M.R. Robilotta, Phys. Rev. C **49** (1994) 1818; Phys. Rev. C **52** (1995) 531.

[39] Th.A. Rijken, Ann. Phys. **208** (1991) 253.

[40] L. Ray, Phys. Rev. C **35** (1987) 1072.

[41] V.G.J. Stoks, R.A.M. Klomp, M.C.M. Rentmeester and J.J. de Swart, Phys. Rev. C **48** (1993) 792.

[42] R.A. Arndt, J.S. Hyslop III and L.D. Roper, Phys. Rev. D **35** (1987) 128; R.A. Arndt and L.D. Roper, Scattering Analysis Interactive Dial-in Program (SAID), Virginia Polytechnic Institute and State University; R.A. Arndt, private communication.

[43] T.E.O. Ericson, Nucl. Phys. **A416** (1984) 281c.

[44] J.A. Nelder and R. Mead, Computer Journal **7** (1965) 308; W.H. Press, B.P. Flannery, S.A. Teukolsky and W.T. Vetterling, *Numerical Recipes – The Art of Scientific Computing* (Cambridge University Press, Cambridge, 1986), pp. 289-293.

[45] R.P. Haddock, R.M. Salter, Jr., M. Zeller, J.B. Czirr, and D.R. Nygren, Phys. Rev. Lett. **14** (1965) 318. The $I = 1$ parameters of our potential were fitted to the *pp* phase shifts. Therefore the predicted $L = 0$ singlet scattering length is comparable to that measured in *nn* scattering, assuming charge symmetry.

[46] H.P. Noyes, Phys. Rev. **130** (1963) 2025.

# Figure Captions

Figure (1) : Tree graphs contributing to the two-nucleon potential (solid lines are nucleons, dashed lines pions).

Figure (2) : One loop graphs contributing to the two-nucleon potential (double lines represent nucleons or isobars). Only one time ordering is shown for each type of graph. In (d) and (e) we only consider those orderings that have at least one pion or one isobar in intermediate states.

Figure (3) : Examples of two-loop graphs that are *not* included in our potential.

Figure (4) : Best fit (solid curves) to the I = 0 $np$ phase shifts and $\epsilon_1$ mixing angle from Ref. [41] assuming a cut-off parameter $\Lambda = 3.90$ fm$^{-1}$. Errors in the phase shifts, where shown, are from Ref. [42].

Figure (5) : Best fit (solid curves) to the I = 1 $pp$ phase shifts and $\epsilon_2$ mixing angle from Ref. [41] assuming a cut-off parameter $\Lambda = 3.90$ fm$^{-1}$. Errors in the phase shifts, where shown, are from Ref. [42].

Figure (6) : Predictions (solid curves) using the $\Lambda = 3.90$ fm$^{-1}$ cut-off and the parameters in Table I in comparison with the Nijmegen phase shift solution [41] for the I = 0 $np$ phase shifts and $\epsilon_1$ mixing angle to 300 MeV. Errors in the phase shifts, where shown, are from Ref. [42].

Figure (7) : Predictions (solid curves) using the $\Lambda = 3.90$ fm$^{-1}$ cut-off and the parameters in Table I in comparison with the Nijmegen phase shift solution [41] for the I = 1 $pp$ phase shifts and $\epsilon_2$ mixing angle to 300 MeV. Errors in the phase shifts, where shown, are from Ref. [42].

Figure (8) : Deuteron $s$-state (upper curve) and $d$-state (lower curve) radial wave functions from the present NN potential using the $\Lambda = 3.90$ fm$^{-1}$ cut-off and the parameters in Table I.

Figure (9) : Radial potentials for the $^1S_0$ partial wave state at 50 MeV using the $\Lambda = 3.90$ fm$^{-1}$ cut-off and the parameters in Table I. The potentials $W^0$, $W^1$, $W^2$ and $V_{eff}$ defined in Eqs. (35) and (36) are indicated by the dashed, dash-dot, dotted and solid curves, respectively. The dash-dot (dotted) curve corresponds to $W^1$/fm ($W^2$/fm$^2$).

Figure (10): Best fits to the I = 0 $np$ phase shifts and $\epsilon_1$ mixing angle from Ref. [41] assuming $\Lambda = 2.50$ fm$^{-1}$ (dashed curves), 3.90 fm$^{-1}$ (solid curves), and 5.00 fm$^{-1}$ (dotted curves). The solid curves here and in Fig. 4 are identical.

Figure (11): Best fits to the I = 1 $pp$ phase shifts and $\epsilon_2$ mixing angle from Ref. [41] assuming $\Lambda = 2.50$ fm$^{-1}$ (dashed curves), 3.90 fm$^{-1}$

(solid curves), and 5.00 fm$^{-1}$ (dotted curves). The solid curves here and in Fig. 5 are identical.